\documentclass[twocolumn,times]{aastex631}
\usepackage{xcolor}
\usepackage{graphicx} 
\usepackage{threeparttable,booktabs}
\usepackage{makecell}

\graphicspath{{./}{Figures/}}
\usepackage[mathscr]{euscript}
\DeclareSymbolFont{rsfs}{U}{rsfs}{m}{n}
\DeclareSymbolFontAlphabet{\mathscrsfs}{rsfs}

\newcommand{\affuofa}{Steward Observatory, University of Arizona, 933 North Cherry Avenue, Tucson, AZ 85721, USA}

\begin{document}

\title{Active Galactic Nuclei in the Green Valley at z$\sim$0.7}

\author[0000-0001-5962-7260]{Charity Woodrum}
\affiliation{\affuofa}

\author[0000-0003-2919-7495]{Christina C. Williams}
\affiliation{NSF’s National Optical-Infrared Astronomy Research Laboratory, 950 North Cherry Avenue, Tucson, AZ 85719, USA}

\author[0000-0002-7893-6170]{Marcia Rieke}
\affiliation{\affuofa}

\author[0000-0003-4565-8239]{Kevin N. Hainline}
\affiliation{\affuofa}

\author[0000-0002-4684-9005]{Raphael E. Hviding}
\affiliation{\affuofa}
\affiliation{Max-Planck-Institut für Astronomie, Königstuhl 17, D-69117 Heidelberg, Germany}

\author[0000-0001-7673-2257]{Zhiyuan Ji}
\affiliation{\affuofa}

\author[0000-0001-5448-1821]{Robert Kennicutt}
\affiliation{\affuofa}
\affiliation{Department of Physics and Astronomy and George P. and Cynthia Woods Mitchell Institute for Fundamental Physics and Astronomy, Texas A\&M University, 4242 TAMU, College Station, TX 77843-4242, USA}

\author[0000-0001-9262-9997]{Christopher N. A. Willmer}
\affiliation{\affuofa}

\begin{abstract}
We present NIR spectroscopy using MMT/MMIRS for a sample of twenty-nine massive galaxies ($\mathrm{log\ M_* / M_{\odot} \gtrsim10}$) at $\mathrm{z\sim0.7}$ with optical spectroscopy from the LEGA-C survey.
Having both optical and NIR spectroscopy at this redshift allows us to measure the full suite of rest-optical strong emission lines, enabling the study of ionization sources and the rest-optical selection of active galactic nuclei (AGN), as well as the measurement of dust-corrected $\mathrm{H\alpha}$-based SFRs.
We find that eleven out of twenty-nine galaxies host AGN.
We infer the nonparametric star formation histories with the SED fitting code \texttt{Prospector} and classify galaxies as star-forming, green valley, or quiescent based on their most recent sSFRs.
We explore the connection between AGN activity and suppressed star formation and find that $89\pm15\%$ of galaxies in the green valley or below host AGN, while only $15\%\pm8\%$ of galaxies above the green valley host AGN. 
We construct the star-forming main sequence (SFMS) and find that the AGN host galaxies are 0.37 dex below the SFMS while galaxies without detectable AGN are consistent with being on the SFMS. However, when compared to a bootstrapped mass-matched sample, the SFRs of our sample of AGN host galaxies are consistent with the full LEGA-C sample.
Based on this mass-matched analysis, we cannot rule out that this suppression of star formation is driven by other processes associated with the higher mass of the AGN sample.
We therefore cannot link the presence of AGN activity to the quenching of star formation.

\end{abstract}


\section{Introduction} \label{sec:intro}
One of the most outstanding open questions in galaxy evolution is identifying the physical processes that drive the quenching of star formation in massive galaxies. A critical finding is the existence of the galaxy bimodality: the majority of galaxies are either blue, lower mass, and star-forming, or are red, massive galaxies that no longer form stars \citep[][]{Strateva2001, Baldry2004, Baldry2006, Bell2004, Willmer2006, Franzetti2007}. The star-forming population follows a well-defined correlation between the star formation rate (SFR) and stellar mass ($\mathrm{M_*}$), known as the star forming main sequence \citep[SFMS;][]{Daddi2007, Noeske2007, Salim2007, Karim2011, Rodighiero2011, Whitaker2012, Whitaker2014, Speagle2014, Renzini2015, Schreiber2015, Tomczak2016, Leslie2020, Leja2022}. 
Between the blue cloud and the red sequence lies a transitional region, often called the green valley \citep{Bell2004, Faber2007, Martin2007, Schiminovich2007, Wyder2007, Mendez2011, Goncalves2012}. Due to its sparse population, it has been suggested that galaxies rapidly \citep[$\mathrm{\approx10^8}$ yr; e.g.][]{Wild2009, Barro2013, Moutard2016, Wu2018, Belli2019, Wild2020} transition through this region, making it a short-lived phase in galactic evolution.
Studying galaxies in the green valley, and any physical processes that preferentially occur there, can provide critical constraints on galaxy quenching \citep[e.g.,][]{Strateva2001, Bell2004, Faber2007}.

One hypothesized process for quenching involves supermassive black holes (SMBHs), which are now understood to co-evolve and grow in mass alongside the mass growth of their host galaxy. Active galactic nucleus (AGN) activity can induce feedback such as winds, jets, and radiation \citep[for reviews, see][]{Alexander2012, Fabian2012}. 
AGN feedback is thought to possibly quench galaxies through many different mechanisms, such as heating the gas or removing gas and dust from the host galaxy \citep[e.g.,][]{Silk1998, Springel2005}. Also, low-power AGN may inject turbulence, which could potentially stabilize the gas disk from fragmentation and prevent star formation \citep{Alatalo2015}. However, simulations show that these mechanisms can also occur as a result of many other processes in the evolution of a galaxy unrelated to AGN activity, such as virial shock heating \citep{Birnboim2003, Keres2005, Dekel2006}, stellar feedback \citep{Hopkins2010, Grudic2019}, morphological quenching \citep{Martig2009, Gensior2020}, and cosmological starvation \citep{Feldmann2015}. Many empirical studies have attempted to place constraints on these processes \citep[e.g.][]{Tremonti2007, Williams2015, Diamond-Stanic2012, Williams2017, Forster2019, Spilker2020, Williams2021, Lammers2022} and have also found evidence for environmental quenching \citep{Peng2010, Peng2015}.
Modern galaxy formation simulations require incorporating feedback into the galaxy formation process to match observational properties of galaxy populations, such as the mass function and colors of quiescent galaxies. At the most massive end, AGN are thought to be the source of this feedback and without it, both cosmological simulations and semi-analytic models produce too many massive blue galaxies \citep{Hatton2003, Croton2006, Gabor2011}. However, the feedback implementations used in these simulations and models vary substantially between each other yet produce relatively similar galaxy populations \citep{Naab2017}. Even though there is strong theoretical evidence for AGN quenching, it has been challenging to causally link AGN to quenching empirically. Therefore, it is vital to place observational constraints on the impact of AGN on their host galaxies.

AGN can be identified across the entire electromagnetic spectrum. For example, the narrow-line regions surrounding AGN are traced by nebular emission lines at rest-frame optical wavelengths. Some of the most widely used AGN selection techniques are diagnostics called the Baldwin, Phillips, and Terlevich diagrams \citep[BPT;][]{Baldwin1981}, which are used to determine the dominant source of ionization in galaxies. This method is one of the most sensitive AGN selection techniques, as it is able to probe to lower accretion rates than other techniques that use X-ray emission or IR colors \citep{Juneau2013}. However, BPT-derived optical selection traditionally requires spectroscopic coverage of [\ion{O}{3}] and H$\mathrm{\beta}$, as well as the $\mathrm{H\alpha}$, [\ion{N}{2}], or [\ion{S}{2}] emission lines, which have longer wavelengths and are redshifted out of the visible spectrum at $\mathrm{z\gtrsim0.5}$. 
Consequently, large statistical studies using BPT diagrams have been limited to low redshift, such as with the Sloan Digital Sky Survey (SDSS). To look for observational evidence of AGN quenching, we need to study galaxies at higher redshifts while they are in the process of quenching. A solution to this problem is follow-up observations with NIR spectrographs. Sky emission is very bright in this part of the electromagnetic spectrum, making NIR observations challenging from ground-based facilities. Therefore, completing the full suite of diagnostics at higher redshifts currently requires significant investments from both optical and NIR spectrographs, such as with the MOSFIRE DEEP Evolution Survey \citep[MOSDEF;][]{Kriek2015} and the Keck Baryonic Structure Survey with MOSFIRE \citep[KBSS-MOSFIRE;][]{Rudie2012, Steidel2014, Strom2017}.

By studying populations of galaxies at different epochs, it has been shown that massive galaxies form rapidly at $\mathrm{z>2}$ \citep[e.g.,][]{Thomas2010, McDermid2015} and most are fully assembled and quenched by $\mathrm{z\approx1}$ \citep{Renzini2006}. We can also explore the \textit{individual} evolutionary pathways of galaxies by utilizing the archaeological record of their stellar populations \citep[e.g.,][]{Thomas1999, Renzini2006, Graves2009, Trager2009, Webb2020, Tacchella2022a, Woodrum2022}. However, as the stellar populations age, these star formation history (SFH) signatures fade. It is therefore necessary to study the SFHs of galaxies closer to the peak epoch of quenching at $\mathrm{z\approx1}$ to provide constraints on their quenching mechanisms.

Traditionally, parametric models, such as the commonly used delayed-$\mathrm{\tau}$ model, have been used to infer the detailed SFHs of galaxies using stellar population modeling. Quenching events may cause sharp transitions in SFR(t), which cannot be captured by these models. A solution to this problem is to infer nonparametric SFHs, which use a highly flexible functional form. Nonparametric models have been shown to match the complexity of galaxy SFHs, such as sharp bursts of star formation or quenching events \citep{Leja2019, Lower2020, Johnson2021}. To infer accurate SFHs of galaxies, both high-quality spectra and photometry are needed to break the SFH-dust-metallicity degeneracy \citep{Tacchella2022a}. Studies with such high quality data have often been limited to the local universe, e.g. SDSS. The Large Early Galaxy Astrophysics Census (LEGA-C) Survey \citep{vanderWel2016, Straatman2018, vanderWel2021} has produced similar quality data for galaxies at $\mathrm{0.6 < z < 1}$ with high-signal-to-noise spectroscopy, enabling the estimation of the properties of galaxies to be characterized in exquisite detail. However, due to their higher redshifts, these galaxy spectra currently lack the full suite of diagnostics available in SDSS, i.e. all of the strong emission lines in the rest-optical ($\mathrm{\lambda_{rest}\approx3000-7000}$\AA). 

To complete this full suite of diagnostics, we follow up a subsample of LEGA-C galaxies to obtain deep NIR spectroscopy using the MMT and Magellan infrared spectrograph \citep[MMIRS;][]{McLeod2012} at the 6.5 m MMT Observatory, targeting three strong rest-frame optical emission lines, $\mathrm{H\alpha}$, [\ion{N}{2}], and [\ion{S}{2}]. By combining these line fluxes with [\ion{O}{3}] and H$\mathrm{\beta}$ from LEGA-C, we are thus able to construct the [\ion{N}{2}] BPT diagram.  We infer the individual, nonparametric SFHs for twenty-nine galaxies at $\mathrm{z\sim0.7}$. We compare the SFHs of galaxies that host AGN to galaxies without detectable AGN to determine if there is a link between the presence of an AGN and the suppression of star formation. We thus explore the role AGN have on the quenching of massive galaxies at $\mathrm{z\sim0.7}$, soon after the peak epoch of quenching \citep[e.g.][]{Bell2004, Faber2007, Muzzin2013a, Tomczak2014, Davidzon2017}.

In Section \ref{sec:data}, we present our data and data reduction methods. In Section \ref{sec:SEDfitting}, we explain our data analysis methods and models. In Section \ref{sec:AGN}, we explain our AGN selection methods. In Section \ref{sec:SFHs}, we discuss the inferred nonparametric SFHs and compare between AGN host galaxies and non-AGN galaxies. In Section \ref{sec:discussion} we compare our results to other studies in the literature. We assume a flat $\Lambda$ cold dark matter ($\Lambda$CDM) cosmology with WMAP-9 parameters, $\Omega_m=0.287$, $H_0=69.3\ \mathrm{km\ s^{-1}\ Mpc^{-1}}$ \citep{Hinshaw2013}.

\section{Data and Observations}\label{sec:data}

\subsection{LEGA-C and UltraVISTA}\label{sec:legac}
LEGA-C \citep{vanderWel2016, Straatman2018, vanderWel2021} is an ESO public spectroscopic survey conducted over 128 nights with VIMOS on the Very Large Telescope in the wide-area COSMOS field \citep{Scoville2007}. The survey includes $\mathrm{\sim 3500}$ $\mathrm{K_s}$-band selected galaxies at redshifts $\mathrm{0.6<z<1.0}$ with stellar masses $\mathrm{M_* > 10^{10} M_{\odot} }$. The 20 hour long integrations produce continuum spectra with S/N $\mathrm{\sim 20 \text{\AA}^{-1}}$ and high-resolution ($\mathrm{\mathcal{R} \approx 3500}$; \citealt{Straatman2018}) at the observed wavelength range $\mathrm{6300 \text{\AA} \leq \lambda \leq 8800 \text{\AA}}$. We use the spectra from the third data release \citep[DR3:][]{vanderWel2021}.

In addition, we utilize the exceptional ancillary data for the LEGA-C galaxies from the UltraVISTA catalog \citep{Muzzin2013a}, which is a collection of photometric data across 30 passbands from 0.2 to 24$\mathrm{\mu m}$. This catalog includes UV imaging from the GALEX satellite \citep{Martin2005}, optical imaging from the Canada–France–Hawaii Telescope (CFHT) and Subaru telescope \citep{Taniguchi2007, Capak2007}, near-infrared data from the VISTA telescope \citep{McCracken2012}, and mid-infrared data from Spitzer \citep{Sanders2007, Frayer2009}. The IRAC and MIPS deblended photometry from Spitzer was measured using a source-fitting code that has been thoroughly tested for many $K_s$ selected catalogs \citep[e.g.][]{Labbe2005, Wuyts2007, Marchesini2009, Williams2009, Whitaker2011, Marchesini2012, Labbe2010, Labbe2013}. For more details, see Section 3.5 of \citet{Muzzin2013a}.

\begin{deluxetable*}{lll}[!htbp]
\tablecaption{\texttt{Prospector} Parameters \label{tab:priors}}
\tablewidth{0pt}
\tablehead{
\colhead{Parameter} & \colhead{Description} & \colhead{Prior}}
\decimalcolnumbers
\startdata
log $M/M_{\odot}$ & Total mass formed & Uniform: min=9.5, max=12.5 \\
$z$ & Redshift & Uniform: min=$z_{spec}-0.005$, max=$z_{spec}+0.005$ \\
log $Z/Z_{\odot}$ & Stellar metallicity & Uniform: min=-1.0, max=0.19 \\
$n$ & \makecell{Power-law modifier to shape of the \citet{Calzetti2000} attenuation \\ curve of the diffuse dust} & Uniform: min=-1.0, max=0.4\\
$\tau_{dust,2}$ & Diffuse dust optical depth & Clipped normal: min=0, max=4, $\mu$=0.3, $\sigma$=1 \\
$\tau_{dust,1}$ & Birth-cloud dust optical depth & Clipped normal in ($\tau_{dust,1}/\tau_{dust,2}$): min=0, max=2, $\mu$=1, $\sigma=0.3$ \\
$\gamma_{e}$ & Mass fraction of dust in high radiation intensity & Log-uniform: min=$10^{-4}$, max=0.1 \\
$U_{min}$ & Minimum starlight intensity to which the dust mass is exposed & Uniform: min=0.1, max=15 \\
$q_{PAH}$ & Percent mass fraction of PAHs in dust & Uniform: min=0, max=7.0 \\
$f_{AGN}$ & AGN luminosity as a fraction of the galaxy bolometric luminosity & Log-uniform: min=$10^{-5}$, max=3 \\
$\tau_{AGN}$ & Optical depth of AGN torus dust & Log-uniform: min=5, max=150 \\
log $\mathrm{Z_{gas}/Z_{\odot}}$ & Gas-phase metallicity & Uniform: min=-2.0, max=0.5 \\
log $\mathrm{U_{gas}}$ & Gas Ionization Parameter & Uniform: min=-4.0, max=-1.0 \\
$\sigma$ & Velocity Dispersion & Uniform: min=$\sigma_{LEGA-C}-50$, max=$\sigma_{LEGA-C}+50$ \\
SFR Ratios & \makecell{Ratio of the SFRs in adjacent bins; $N_{SFH}$-1 bins total with  \\ $N_{SFH}$=10} & Student’s t-distribution: $\sigma = 0.3$, $\nu = 2$ \\
$\mathrm{f_{out}}$ & Spectra outlier fraction & Uniform: min=$10^{-5}$, max=0.1 \\
$\mathrm{j_{spec}}$ & Spectral noise multiplier & Uniform: min=0, max=3.0\\
\enddata
\tablecomments{$z_{spec}$ and $\sigma$ are the spectroscopic redshift and velocity dispersion from LEGA-C DR3}
\end{deluxetable*}

\subsection{Sample Selection and Follow-up at MMT}\label{sec:mmt}
In this work, we select a subsample of galaxies drawn from the LEGA-C survey. To complete the full suite of rest-frame diagnostics, we obtained follow-up NIR observations using the MMIRS spectrograph \citep{McLeod2012} at MMT Observatory, targeting three strong rest-frame optical emission lines, including $\mathrm{H\alpha}$, [\ion{N}{2}], and [\ion{S}{2}]. For our selection, we prioritized LEGA-C galaxies for which the measured line fluxes for H$\mathrm{\beta}$ and [\ion{O}{3}] were well-detected with S/N$\mathrm{>3}$ after continuum subtraction and were not likely to be contaminated by sky lines. We chose mask pointings that would maximize the number of these galaxies for each mask. When remaining mask space was available, we assigned slits to other LEGA-C galaxies as fillers if they had a quality flag in the catalog that indicated they were primary survey targets that can be used for scientific purposes (see Section 3.3 of \citet{Straatman2018} for more details). We obtained spectra with four masks for a total of sixty-one galaxies, thirty-five filler galaxies and twenty-six galaxies that met our primary criteria. 
Twenty-nine galaxies in total showed H$\mathrm{\alpha}$ emission and are the sample discussed in this paper, seven of which were filler galaxies. 

Our NIR spectroscopy is observed with MMIRS, which is an IR multi-object spectrograph with a wide field of view ($\mathrm{4\arcmin \times 7\arcmin}$). We observed with the J grism and the zJ filter with a wavelength range of 0.94-1.51$\mu m$ and $\mathrm{\mathcal{R}\approx1020}$ at the mid-wavelength of 1.225$\mathrm{\mu m}$. We obtained spectra with a total of 4 masks, each of which was observed for $\sim$5.5-6.5 hours (combining individual exposures of 300 s), with average seeing of 0.51-1.08$\arcsec$. The slit widths are 1$\arcsec$ and the slit lengths are 8$\arcsec$ to be consistent with the LEGA-C observations. The orientation of the slits are shown for our observations as well as the LEGA-C observations in Appendix \ref{sec:appendix}.

To investigate selection effects related to the local environment of the galaxies in our sample, we use the overdensity value estimates from \citet{Darvish2016}. The overdensity is defined as the local number density of galaxies, which is estimated using a Voronoi tessellation technique. The mean and standard deviation of the overdensity value for the full LEGA-C sample is $\mathrm{log(1 + \delta) =0.25+-0.28}$. For the sample studied in this paper, the mean and standard deviation of the overdensity value is $\mathrm{log(1 + \delta) =0.03+-0.31}$. Therefore, our sample is not in systematically overdense environments compared to the full LEGA-C sample or the field overall. We therefore find no evidence that potential quenching mechanisms are driven by the local environmental density of the galaxies in our sample.

\subsection{Data Reduction}
The raw frames were processed through the MMIRS data reduction pipeline \citep[][]{Chilingarian2015}. The raw data consist of science images and dark exposures of the same duration as the science data. All data use a non-destructive readout such that a 3 dimensional file is created where the third dimension contains the individual readouts. The dark frames are averaged on a frame-by-frame basis and then subtracted from each of the science frames. Subsequently, data rate frames are calculated by using a linear fit for each pixel which also allows to reject cosmic rays. These data rate frames are calibrated in wavelength using the OH sky lines and re-sampled to be linear in wavelengths. The linearized spectra are combined on a per-mask basis taking into account the dither pattern used during the observations, to generate the combined 2-dimensional spectrum.

We extract 1D spectra by first fitting the profile along the spatial axis with a Gaussian function and then using optimal extraction \citep[][]{Horne1986}. The aperture size for this extraction varies from galaxy to galaxy, but has an average FWHM of 1.2\arcsec. The LEGA-C 1d spectra were extracted with an aperture that also varied from galaxy to galaxy, however the individual aperture sizes are not published, but they have a very similar typical size of 1\arcsec. The absolute flux calibration was done using the photometry from UltraVISTA, as was done for the LEGA-C spectra. To scale the MMT spectra to the photometry, we first fit the photometry alone with \texttt{Prospector} \citep[e.g.,][see Section \ref{sec:prospector}]{Johnson2021} where the fit redshift is set to the spectroscopic redshift to obtain a best-fit photometric SED. We use a median filter, 100\AA\ wide, along the wavelength range for both the MMT spectra and the best-fit photometric SED and find the ratio between the two. We use this ratio to scale the MMT spectra to the best-fit photometric SED in a wavelength-dependent way. We note that any differential wavelength effects should be negligible in our AGN selection technique, which depends on the intensity ratios of emission lines that are sufficiently close in wavelength. We also note that while the flux calibration does compensate for slit losses under the assumption that the spectrum is the same shape across the galaxy, it does not account for any spatial gradients along the slit that may vary depending on wavelength.

\section{Data Analysis}\label{sec:SEDfitting}
We use SED fitting to infer physical properties for our sample including $\mathrm{M_*}$, the detailed SFHs, and dust attenuation. In addition, we fit emission lines to measure $\mathrm{H\beta}$, [\ion{O}{3}], [\ion{N}{2}], $\mathrm{H\alpha}$, and [\ion{S}{2}] fluxes. In this section we discuss the details of our analysis. The simultaneous fitting of the photometry and LEGA-C optical spectra with \texttt{Prospector} is described in Section \ref{sec:prospector}. The emission line fitting of the spectra with the flexible framework GELATO \citep{Hviding2022} is described in Section \ref{sec:gelato}.

\subsection{\texttt{Prospector}}\label{sec:prospector}
We simultaneously fit the photometry and LEGA-C optical spectra with the \texttt{Prospector} inference framework. \texttt{Prospector} \citep{Johnson2021} uses the Flexible Stellar Population Synthesis code (\texttt{FSPS}) \citep{Conroy2009} via \texttt{python-FSPS} \citep{ForemanMackey2014}. The posterior distributions are sampled using the dynamic nested sampling code \texttt{dynesty} \citep{dynesty:2020}. 

We use a similar physical model as in \citet{Woodrum2022}, modified for star-forming galaxies. In brief, we employ the MIST stellar evolutionary tracks and isochrones \citep{Choi2016, Dotter2016} which utilizes the MESA stellar evolution package \citep{Paxton2011, Paxton2013, Paxton2015, Paxton2018}. We use MILES for the stellar spectral library \citep{Vazdekis2015, Falcon-Barroso2011} and adopt a Chabrier IMF \citep{Chabrier2003}. The IGM absorption is modeled after \citet{Madau1995}. For dust attenuation, we assume a flexible attenuation curve with the UV bump tied to the slope of the curve \citep{Kriek2013} and a two-component dust model \citep{Charlot2000}. For nebular emission, we use nebular marginalization, which models the emission lines with a Gaussian fit, therefore decoupling the emission lines from the SFH, see Section \ref{sec:SFHs_measurements}. Compared to our model in \citet{Woodrum2022}, in this work we allowed a wider range in priors for the gas-phase metallicity and gas ionization parameter. Our dust emission model assumes energy balance, where all starlight is attenuated by dust and re-emitted in the IR \citep{daCunha2008}. We use the \citet{Draine2007} dust emission templates, which constrains the shape of the IR SED. We include both noise and calibration models for the simultaneous fitting of both high resolution spectra and photometry. We adopt a ten-component nonparametric SFH using the continuity prior. 

We adopt AGN templates from \citet{Nenkova2008a} and \citet{Nenkova2008b}, which are CLUMPY AGN models incorporated into \texttt{FSPS}. Only dust emission from the AGN's dusty torus is included in this model. The UV and optical emission from the central engine is mostly obscured by the AGN dust torus, and if any emission leaks out it is then attenuated by the galaxy dust attenuation model. CLUMPY AGN models successfully reproduce the observed MIR characteristics of AGN in the local universe \citep[e.g.,][]{Mor2009, Honig2010}, however are not the best method to identify emission from unobscured AGN \citep[see also,][]{Leja2018}. The free parameters for our physical model are listed in Table \ref{tab:priors}.

\subsection{GELATO}\label{sec:gelato}
To study the source of ionization in our sample, we fit the emission lines in the NIR and optical spectra with the flexible framework GELATO \citep{Hviding2022}. GELATO models the stellar continuum as a combination of simple stellar populations (SSPs) from the Extended MILES stellar library \citep[E-MILES;][]{Vazdekis2016}. The SSP models assume a Chabrier IMF and isochrones of \citet{Girardi2000} with solar alpha abundance, and span a range of metallicities and ages.
We extract fluxes for $\mathrm{H\beta}$, [\ion{O}{3}], [\ion{N}{2}], $\mathrm{H\alpha}$, and [\ion{S}{2}]. Each emission line is treated as a Gaussian model and parameterized by its redshift, velocity dispersion (the standard deviation of the Gaussian), and flux. For more information, see \citet{Hviding2022a}. GELATO also allows emission lines to be fit with an additional broad line component. However, fitting with broad lines for our sample does not statistically improve the fit. The GELATO fits for our entire sample are shown and the fluxes are listed in Table \ref{tab:measurements} in Appendix \ref{sec:appendix}.

\begin{figure}[!htbp]
\includegraphics[width=0.5\textwidth]{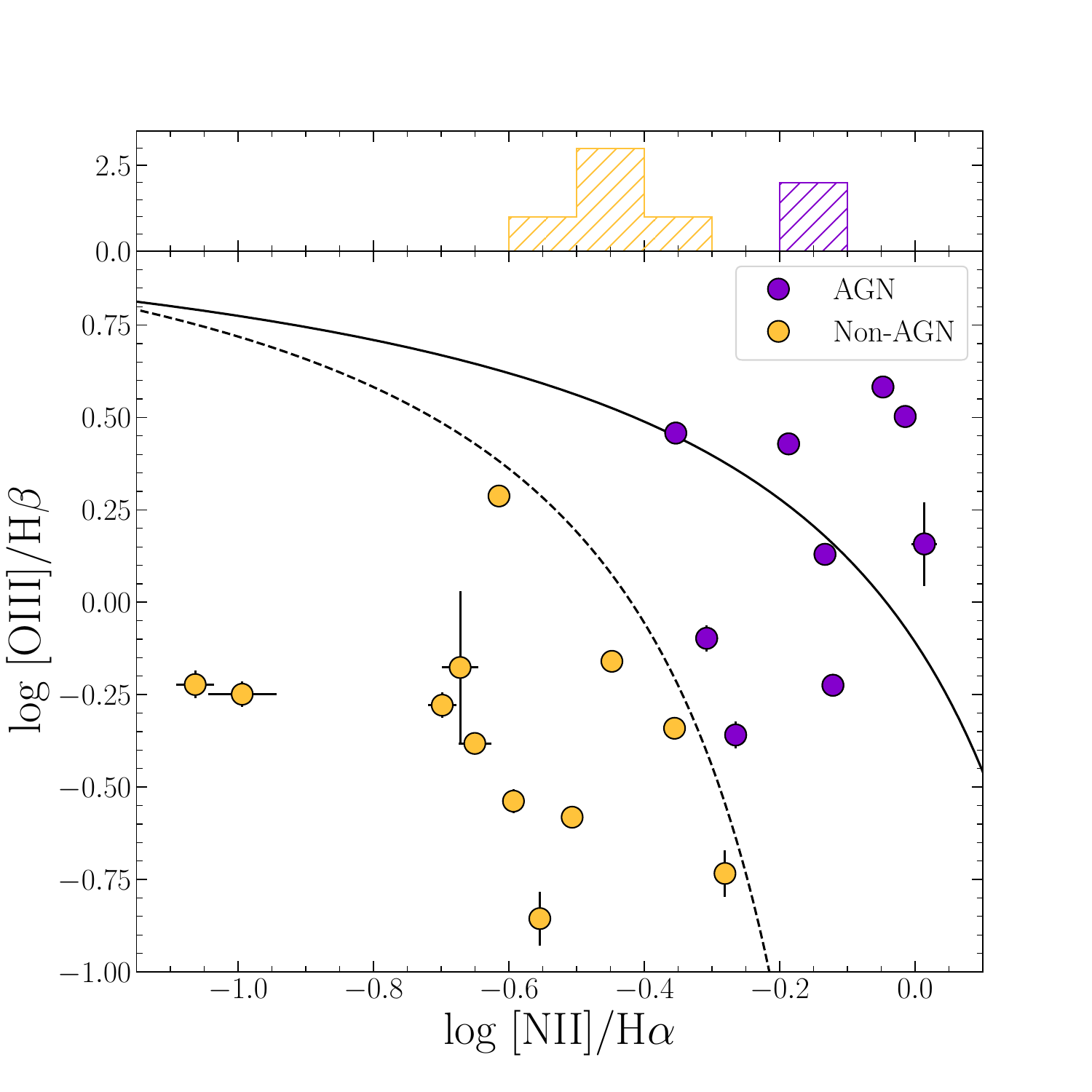}
\caption{[\ion{N}{2}] BPT diagnostic diagram showing the relation between two line ratios: log([\ion{O}{3}]$\lambda5007/H\beta$) vs. log([\ion{N}{2}]$\lambda6584/H\alpha$). The solid curve is from \citet{Kewley2001} and the dashed line is from \citet{Kauffmann2003}. Seven galaxies do not have measurements for [\ion{O}{3}] because the line was redshifted out of range for the LEGA-C observations and are therefore not included in the BPT diagram, however their values for log([\ion{N}{2}]$\lambda6584/H\alpha$) are shown in the top panel as a histogram. Based on this diagram, eleven galaxies in our sample are AGN. \label{fig:bpt}}
\end{figure}

\section{Identifying AGN}\label{sec:AGN}
We identify AGN activity in our sample through the BPT diagnostic diagram. In Figure \ref{fig:bpt}, we show the [\ion{N}{2}] BPT diagram, which compares the ratio of [\ion{O}{3}]$\mathrm{\lambda}$5007\AA\ to H$\mathrm{\beta}$ with [\ion{N}{2}]$\mathrm{\lambda}$6583\AA\ to H$\mathrm{\alpha}$. We note that these line ratios have been shown to evolve at higher redshift \citep[e.g.,][]{Shapley2005, Juneau2014, Steidel2014, Shapley2015}, however galaxies at our redshift of $\mathrm{z\approx 0.7}$ have interstellar medium conditions consistent with local galaxies \citep[e.g.,][]{Kewley2013, Helton2022}. Demarcation lines from \citet{Kewley2001} and \citet{Kauffmann2003} are shown as solid and dashed curves, respectively, to separate star-forming galaxies from AGN. AGN tend to lie above the \citet{Kewley2001} line, while galaxies below this line but above the \citet{Kauffmann2003} line are considered to be in an area on the BPT diagram where composite galaxies lie, meaning both star formation and AGN activity may contribute to the ionization. In this work, we consider composite galaxies to be AGN.

Using this diagnostic, we find that five galaxies have line ratios consistent with AGN ionization, and four are in the composite region. Seven galaxies do not have measurements for [\ion{O}{3}] or $\mathrm{H\beta}$ because the line was redshifted out of range for the LEGA-C observations but not redshifted into the range accessible to MMIRS, lying between the two. These galaxies are therefore not included in the BPT diagram, however their values for log([\ion{N}{2}]$\mathrm{\lambda6584/H\alpha}$) are shown in the top panel as a histogram; we consider two of these galaxies with log([\ion{N}{2}])$>-0.2$ to be either AGN or composite galaxies. Therefore, in total, our sample contains eleven AGN selected using the [\ion{N}{2}] BPT diagram. We note that when using optical diagnostics, it is difficult to separate line emission between excitation sources if shock emission is suspected \citep[e.g.,][]{Kewley2019}. We therefore cannot rule out that the galaxies in the AGN and composite regions may be indicative of the presence of shocks instead of AGN.

We match our sample of galaxies with the Chandra COSMOS Legacy Survey catalogs \citep{Civano2016, Marchesi2016}. Four galaxies in our sample were X-ray detected, all of which were also identified as AGN or composite galaxies with the BPT diagram. We include the 0.5–7.0 keV rest frame luminosities ($\mathrm{L_{X,int}}$) in Table \ref{tab:AGN}. The criterion to be classified as an X-ray AGN requires $\mathrm{L_{X,int} \geq 3\times 10^{42}\ erg\ s^{-1}}$ \citep{Luo2017}. Three of the four X-ray detections meet this criterion.

In addition, we check the IRAC colors for evidence of obscured AGN that might be missed by X-ray or BPT selection. 
We use the IRAC color-color diagram from \citet{Donley2012} and find that none of the galaxies in our sample meet the criteria for IR AGN. This is not surprising because some optically-selected AGN can have IR colors consistent with star-forming galaxies and IR selection is very dependent on AGN to host luminosity ratio \citep{Hviding2022}. In addition, \citet{Leja2018} found that only 46\% of AGN identified with \texttt{Prospector}, defined as $\mathrm{f_{AGN}>0.1}$, would be detected with typical MIR color selections. The two galaxies in our sample with $\mathrm{f_{AGN}>0.1}$ were also identified as AGN or composite using the BPT diagram. 

Some AGN can produce jets which emit synchrotron radiation that can be detected at radio wavelengths. We match our sample with the radio-loud galaxy catalog in \citet{Barisic2017}. They classify radio-loud AGN by using the radio luminosity limit from \citet{Best2005}. We find that one of our galaxies is included in their radio-loud AGN catalog. Based on its location in the BPT diagram, this radio-loud AGN is also a composite galaxy. 

In summary, all of the galaxies in our sample that were classified as AGN from radio, IR, or X-ray emission were also classified as AGN by the BPT diagram. Seven of the BPT-selected AGN do not have detectable X-ray or radio emission. \citet{Juneau2013} also found that optically-selected AGN were not always detected using alternate methods (X-ray, IR, radio) and explained that this is likely because optical emission lines are one of the most sensitive AGN tracers. All of the results for this section are summarized in Table \ref{tab:AGN}.

\begin{figure}[!htbp]
\includegraphics[width=0.5\textwidth]{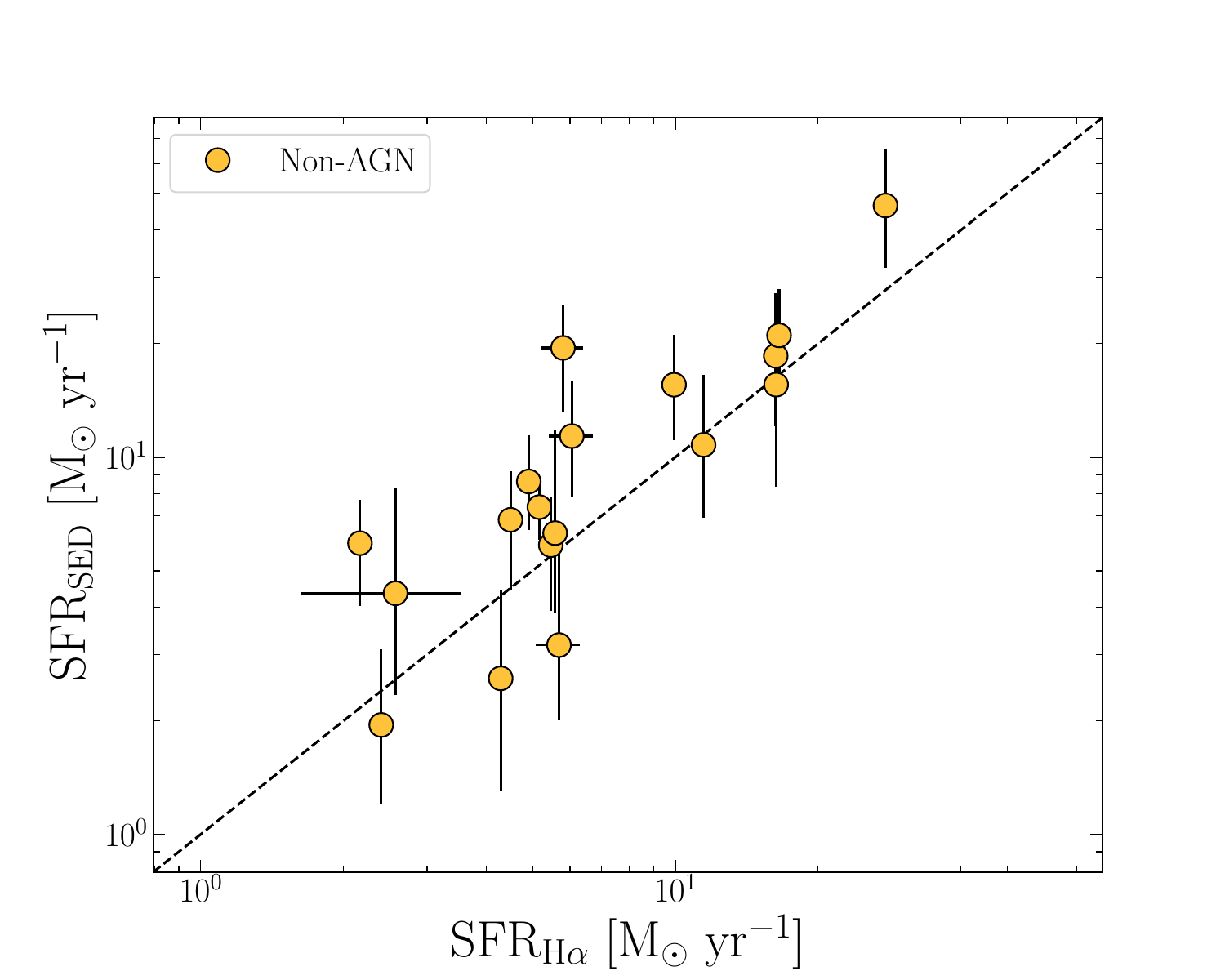}
\caption{Comparison of the dust-corrected SFRs derived from $\mathrm{H\alpha}$ fluxes with the SFRs inferred from the \texttt{Prospector} fits, showing good agreement with only 0.1 dex in scatter. Only galaxies without detectable AGN are shown in this comparison because AGN significantly contribute to H$\mathrm{\alpha}$ flux.} \label{fig:SFR_comparison}
\end{figure} 

\begin{figure*}[!htbp]
\includegraphics[trim={7cm 1cm 6cm 3cm}, clip, width=\textwidth]{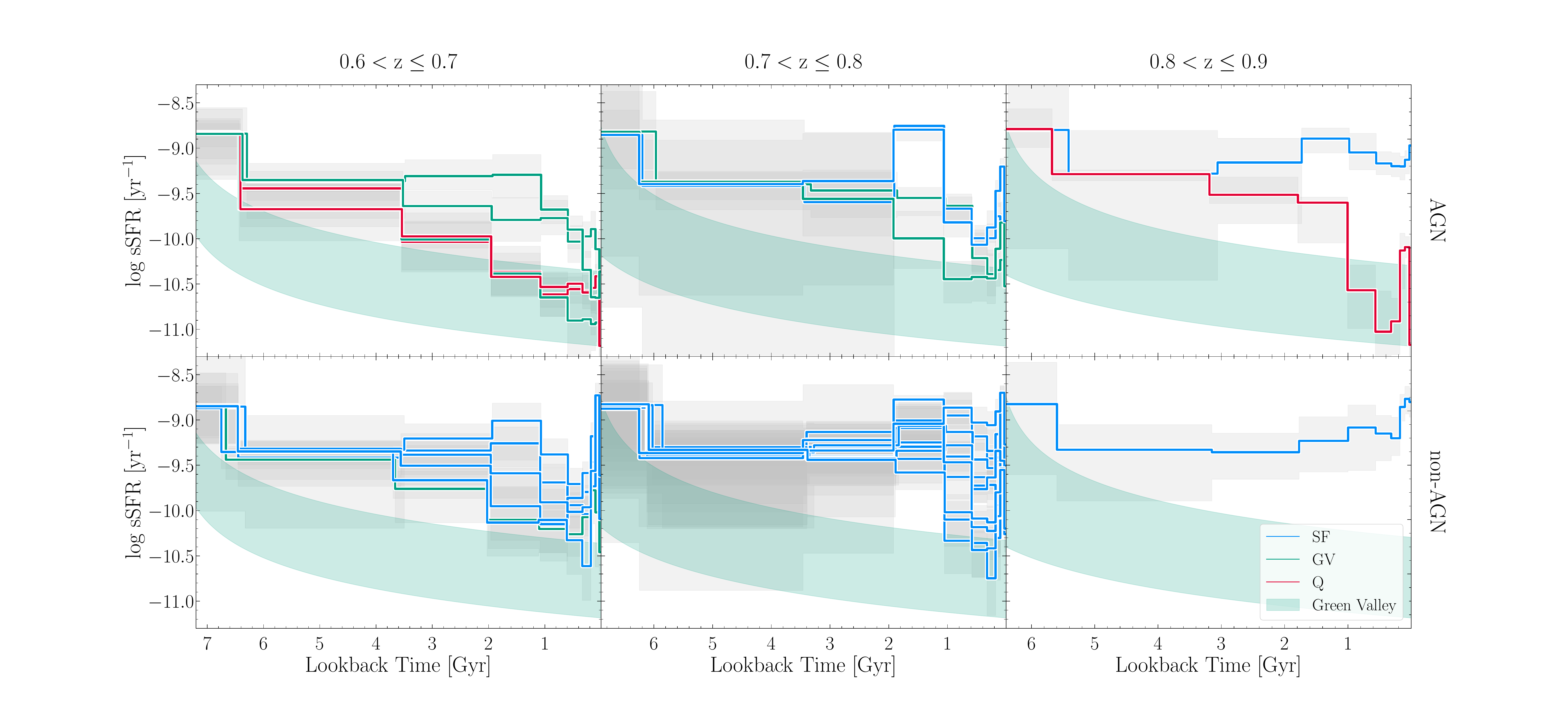}
\caption{The individual SFHs for all galaxies in our sample. The shaded regions show the 16th and 84th percentile range of the posterior probability while the solid lines show the median. The top row shows AGN and the bottom row shows galaxies without detectable AGN. The columns are separated by bins in redshift. The empirically determined green valley is shown as a shaded green region. The individual SFHs are color-coded based on whether galaxies are classified as star-forming (blue), transitioning (green), or quiescent (red) at the epoch of observation. Twenty-two galaxies are star-forming, six galaxies are transitioning, and one galaxy is quiescent. \label{fig:full_SFHs}}
\end{figure*}

\section{Star Formation Histories}\label{sec:SFHs}
\subsection{SFR Measurements}\label{sec:SFHs_measurements}
We aim to measure robust SFRs in order to explore the role of AGN in the quenching of massive galaxies. Many AGN indicators are also SFR indicators, making it difficult to disentangle the two. If AGN are present, they may contribute to the flux of the emission lines. Therefore, for the \texttt{Prospector} SED fitting, we use nebular marginalization, which decouples the emission lines from the SFR, as discussed in Section \ref{sec:SEDfitting}.

Hydrogen recombination lines are thought to be one of the most robust tracers of SFR. In particular, $\mathrm{H\alpha}$ is the SFR indicator of choice, which probes the SFR over the past 10 Myr \citep{Kennicutt2012}. However, as mentioned in Section \ref{sec:intro}, $\mathrm{H\alpha}$ redshifts out of the visible spectrum at $\mathrm{z>0.5}$ and consequently requires NIR spectroscopy. Our NIR data cover the $\mathrm{H\alpha}$ line, enabling a cross-comparison of SFR tracers between the spectrophotometric SED-modeling of the optical data while accurately correcting for dust attenuation. \citet{Tacchella2022a} showed that simultaneously fitting photometry and spectroscopy is needed to break the dust-age-metallicity degeneracy, and that the dust attenuation is mainly constrained by the photometry.

To measure $\mathrm{SFR_{H\alpha}}$, we first correct the $\mathrm{H\alpha}$ flux for dust attenuation using the best inferred dust attenuation law parameters from our \texttt{Prospector} fits. We use the dust-corrected fluxes to derive SFRs using the conversion factor in \citet{Kennicutt2012}. To check if the recent SFR inferred from our SED fitting with \texttt{Prospector} is robust, in Figure \ref{fig:SFR_comparison} we show the $\mathrm{SFR_{H\alpha}}$ compared with the $\mathrm{SFR_{SED}}$. We only include galaxies with no detectable AGN in this comparison because AGN contribute significantly to $\mathrm{H\alpha}$ flux.  We calculate the scatter as the standard deviation of the perpendicular distances to the one-to-one line and find that it is 0.1 dex. Therefore, the results are in excellent agreement.

In Figure \ref{fig:full_SFHs}, we show the SFHs of individual galaxies inferred with our nonparametric model, displaying a diversity of evolutionary pathways for our sample of galaxies. The top 3 panels show galaxies hosting AGN and the bottom 3 panels show galaxies without detectable AGN. The columns represent different redshift bins. The green shaded region is the transition regime, or the ``green valley," defined by \citet{Tacchella2022a} using cuts in specific SFRs (sSFR = SFR/$\mathrm{M_*}$) based on the mass doubling number, which is a measure of the number of times the stellar mass would double over the age of the universe at the current sSFR. \citet{Tacchella2022a} verified these cuts empirically using the SFR-$\mathrm{M_*}$ plane from \citet{Leja2022}. Galaxies with sSFRs below this region are classified as quiescent and galaxies with sSFRs above this region are classified as star-forming.

Using this SFH classification scheme, three galaxies are quiescent, six galaxies are in the green valley, and the remaining twenty galaxies are star-forming. All of the quiescent and green valley galaxies have higher masses ($\mathrm{log\ M_*>10.7\ M_{\odot}}$). Next, we discuss the SFHs of AGN compared to galaxies without detectable AGN.

\begin{deluxetable*}{lllllllc}[!htbp]
\tablecaption{AGN \label{tab:AGN}}
\tablewidth{0pt}
\tablehead{
\colhead{\#} & \colhead{ID} & \colhead{BPT} &  \colhead{$\mathrm{L_{X,int}}$} & \colhead{Radio-loud} & \colhead{$\mathrm{f_{AGN}}$} & \colhead{Galaxy classification} }
\startdata
1 & 151258 & AGN & no & no & 0.0 & GV \\
2 & 254580 & AGN & no & no & 0.1 & GV \\
3 & 165155 & AGN & 3.3E+42 & no & 0.0 & GV \\
4 & 240905 & AGN & no & no & 0.1 & Q \\
5 & 144632 & AGN & 1.6E+42 & no & 0.2 & GV \\
6 & 161993 & Composite & no & no & 0.1 & SF \\
7 & 234067 & Composite & 5.2E+42 & yes & 0.0 & SF\\
8 & 166309 & Composite & 3.2E+42 & no & 0.0 & GV \\
9 & 240675 & Composite & no & no & 0.0 & Q \\
10 & 143863 & AGN/Composite & no & no & 0.3 & SF \\
11 & 162906 & AGN/Composite & no & no & 0.1 & Q \\
\enddata
\tablecomments{All of the galaxies in our sample with detectable AGN are included in this table. AGN that were classified from radio, IR, or X-ray emission were also classified as AGN by the BPT diagram. Column 3 indicates whether the galaxy lies above the \citet{Kewley2001} or \citet{Kauffmann2003} line in the BPT diagram, labeled as AGN or composite, respectively, see Figure \ref{fig:bpt}. Galaxies labelled as AGN/Composite do not have [\ion{O}{3}] measurements because the line was redshifted out of range for the LEGA-C observations, however their values for log([\ion{N}{2}]/$\mathrm{H\alpha}$) are high enough to consider them either AGN or composite galaxies. Column 4 indicates whether the galaxy was detected in the Chandra Legacy Survey \citep[if yes, the luminosities are listed;][]{Civano2016, Marchesi2016}. Column 5 indicates whether the galaxy was determined a radio-loud AGN \citep{Barisic2017}. Column 6 shows the inferred AGN luminosity as a fraction of the galaxy's bolometric luminosity from our \texttt{Prospector} fits. Column 7 indicates whether the galaxy is classified as star-forming (SF), transitioning in the green valley (GV), or quiescent (Q) based on the cuts in sSFR from \citep{Tacchella2022a}.}
\end{deluxetable*}

\subsection{Star Formation Histories of AGN vs.\ non-AGN}\label{sec:sfhs_agn}
\begin{figure*}[!htbp]
\includegraphics[width=\textwidth]{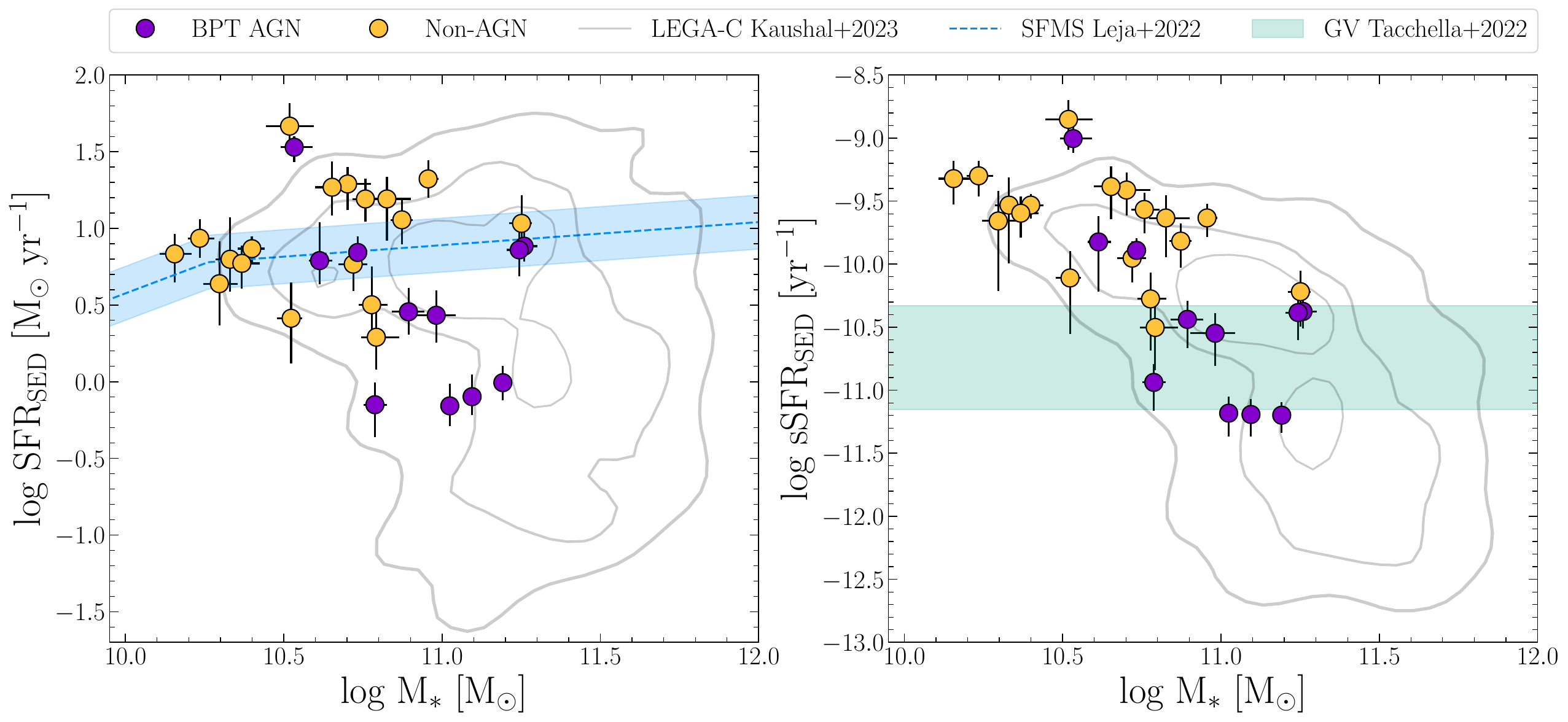}
\caption{The left panel shows the SFR determined from the SED-fitting vs. the stellar mass for the purple and yellow circles. The contours show the 25th, 50th and 75th percentiles from the full LEGA-C sample fit with a similar \texttt{Prospector} model in \citet{Kaushal2023}. The SFMS from \citet{Leja2022} is shown as a blue dashed lines, with 1.5$\times$ above and below the SFMS shown as a shaded region. The right panel shows the sSFR determined from the SED fitting vs.\ the stellar mass. The green valley, as discussed in Section \ref{sec:SFHs}, is shown as a green shaded region. \label{fig:sfms}}
\end{figure*}

The left panel of Figure \ref{fig:sfms} shows the SFMS for the galaxies in our sample using $\mathrm{SFR_{SED}}$, with galaxies hosting BPT AGN shown in purple and galaxies without detectable AGN shown in yellow. \citet{Kaushal2023} fit the entire LEGA-C sample using \texttt{Prospector} with a similar physical model as used in this work. Their results are shown as contours for the 25th, 50th, and 75th percentiles. We compare $\mathrm{SFR_{SED}}$ relative to the redshift-dependent SFMS from \citet{Leja2022}. We calculate the mean and standard error of the mean for $\mathrm{log\ (SFR_{SED}/SFR_{MS}(z))}$ and find that for AGN host galaxies it is $\mathrm{-0.37\pm0.15}$ and for galaxies without detectable AGN it is $\mathrm{0.14\pm0.07}$. Therefore AGN host galaxies lie significantly below the SFMS while galaxies without detectable AGN are consistent with being on or above the SFMS.

The right panel of Figure \ref{fig:sfms} shows the sSFR vs.\ stellar mass, as determined from the SED fitting, with the green valley at $\mathrm{z=0.75}$ shown as a green shaded region. As mentioned in Section \ref{sec:SFHs}, the majority of galaxies in our sample are star-forming (above the green valley), based on the inferred SFHs from the SED fitting with \texttt{Prospector}. Six galaxies are in the transition region, five of which are AGN. In addition, the three recently quenched galaxies, which lie below the green valley, are AGN. Therefore, $89\pm15\%$ of galaxies in the green valley or below host AGN, while $15\pm8\%$ of galaxies above the green valley host AGN. The uncertainties for these percentages were calculated using the binomial distribution. In Section \ref{sec:sfhs_agns_mass} we discuss if this suppression of star formation in AGN host galaxies is consistent with the overall trend with increasing stellar mass of the parent sample.

Using the nonparametric SFHs, we calculate the amount of time spent in the green valley ($\mathrm{t_{GV}}$) for the five AGN, which differs significantly between galaxy and is in the range $\mathrm{t_{GV}\approx 0.03-2}$ Gyr. This is consistent with the large diversity of quenching timescales for quiescent galaxies, defined as time spent in the green valley before quiescence, found by \citet{Tacchella2022a}, whose definition of green valley is used in this work.

\subsection{Star Formation History Comparison as a Function of Stellar Mass}\label{sec:sfhs_agns_mass}
The number of AGN in our sample increases with $\mathrm{M_*}$, in agreement with many other studies \citep{Kauffmann2003, Xue2010, Aird2012, Ji2022}. This is not surprising given the positive correlations between the black hole mass ($\mathrm{M_{BH}}$), bulge mass, and stellar mass \citep{Kormendy2013, Delvecchio2019}. Specifically, galaxies with larger $\mathrm{M_*}$ also have larger $\mathrm{M_{BH}}$ with higher absolute accretion rates \citep{Mullaney2012, Yang2018}. Therefore, it is expected that AGN are more likely to be detected in more massive galaxies. In addition, the majority of the high mass galaxies in the full LEGA-C sample are below the SFMS. As a result, it is necessary to compare SFRs as a function of $\mathrm{M_*}$ when looking for differences between AGN host galaxies and galaxies without detectable AGN to determine if any SFR differences are driven by the AGN and not other physical processes associated with higher stellar masses.

To accomplish this, we need to compare the star formation properties of AGN host galaxies to a mass-matched sample of non-AGN galaxies. 
Therefore, we compare the SFRs of our AGN host galaxy sample with a mass-matched bootstrapped sample from the \cite{Kaushal2023} catalog. We find that the mean and standard deviation of the $\mathrm{SFR_{SED}}$ for our AGN host galaxies is $\mathrm{2.9\pm9.1}$ and for the mass-matched bootstrapped sample it is $\mathrm{3.4\pm3.1}$. These are consistent with one another, meaning that the SFRs of our AGN host galaxies follow the same SFR trends at a given stellar mass as the full LEGA-C sample. We therefore cannot connect the lower SFRs with the current AGN activity, as it is possibly driven by other processes associated with the higher mass of the AGN sample.

\section{Discussion}\label{sec:discussion}
In this work, we explore the connection between AGN activity and star formation. Many studies have compared the SFRs of AGN host galaxies to galaxies without detectable AGN and find varying results. We find that the green valley galaxies in our sample preferentially host AGN, consistent with other studies \citep{Nandra2007, Martin2007, Salim2007, Schiminovich2007, Hasinger2008, Silverman2008, Cimatti2013, Man2019, Dodd2021, Lammers2022}. Compared to galaxies without detectable AGN, some studies find higher SFRs for the AGN host galaxies \citep{Silverman2009, Rovilos2012, Santini2012b, Juneau2013, Rosario2015, Magliocchetti2016}, others find similar SFRs \citep{Harrison2012, Rosario2013, Stanley2015, Xu2015}, while others find lower SFRs \citep{Salim2007, Mullaney2012, Page2012, Hardcastle2013, Barger2015, Gurkan2015, Shimizu2015, Leslie2016, Aird2019}. 

\begin{figure}[!htbp]
\includegraphics[width=0.45\textwidth]{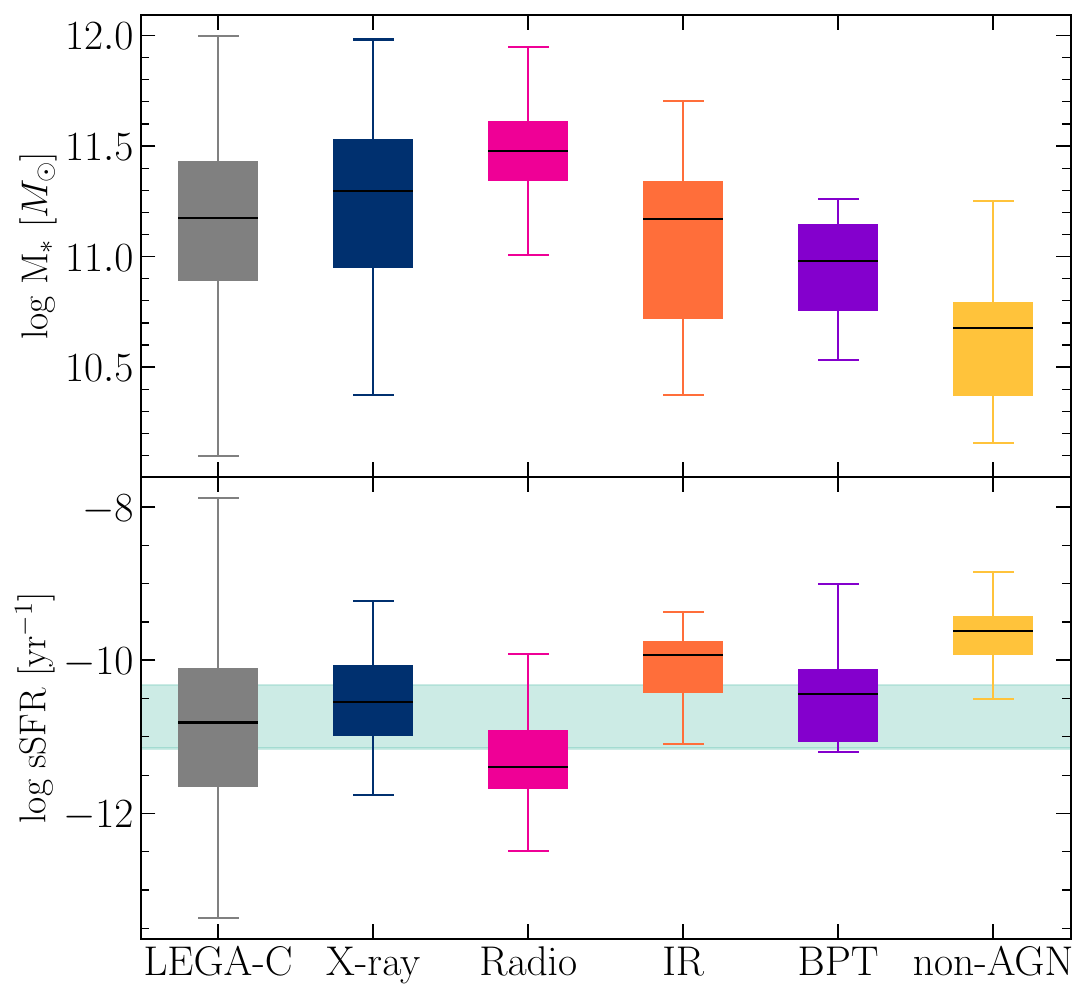}
\caption{The distributions of the stellar masses and sSFRs inferred for the entire LEGA-C sample \citep{Kaushal2023}, with columns showing the X-ray AGN, the radio AGN \citep{Barisic2017}, the IR AGN, the BPT AGN, and the galaxies in our NIR sample that were not classified as AGN on the BPT diagram. The green valley, at z=0.75, is shown as a green shaded region. \label{fig:mass_sfr_dist}}
\end{figure}

To compare to these results, we utilize the entire LEGA-C sample from the \citet{Kaushal2023} \texttt{Prospector} catalog which uses a similar physical model as in this work. We select X-ray, IR, and radio AGN in the full LEGA-C sample using the same selection techniques discussed in Section \ref{sec:AGN}. Figure \ref{fig:mass_sfr_dist} shows the distributions of the stellar masses and sSFRS for each AGN selection type. Consistent with other results, the IR-selected AGN tend to have enhanced star formation \citep{Ellison2016, Ji2022}, while X-ray AGN have a broad range of sSFRs, showing both enhanced and suppressed star formation \citep{Juneau2013, Aird2019, Ji2022}. The radio-selected AGN tend to have low sSFRs and high stellar masses.

These differences in the SFRs of AGN host galaxies may be due to the AGN selection technique \citep{Ji2022} or the type of AGN \citep{Ellison2016, Comerford2020, Georgantopoulos2023}. For example, \citet{Ellison2016} suggest that IR-selected AGN may have a higher fraction of mergers compared to the merger fraction of optically selected AGN, and these mergers are also thought to trigger episodes of star formation. Other studies have found a connection between radio-loud AGN and quiescence \citep{Best2014, Barisic2017}. \citet{Trump2015} show that BPT-selected AGN are preferentially found in high-mass, green, moderately star-forming hosts due to ``star formation dilution," which causes a significant bias against BPT AGN selection in lower mass star-forming galaxies. Therefore, different techniques may select for AGN and host galaxies in different stages of their co-evolutionary pathways, or they may simply be the result of observational selection bias. 

Galaxies can reside in the green valley for two main reasons: (1) a quiescent galaxy can move from the red sequence to the green valley via the rejuvenation of star formation or (2) the galaxy is in the process of quenching. Other studies that find lower SFRs for AGN host galaxies have suggested the former scenario, where the build-up of the central bulge through inflowing gas can trigger AGN activity as well as rejuvenate star formation \citep{Hasinger2008, Aird2019}. With our detailed individual, non-parametric SFH modeling that has been shown to capture rejuvenation events \citep[e.g.,][]{Tacchella2022a, Woodrum2022}, we are able to test this scenario for the first time. We use the following criteria to define rejuvenation: galaxies that were once quiescent (their median sSFR fell below the green valley) and have risen to either the green valley or above at some time after. We inspect the individual SFHs of the green-valley AGN host galaxies in our sample and find that they do not meet this rejuvenation criteria. We also inspect the SFHs of the rest of our sample and find that none of our sample meets this rejuvenation criteria. Therefore, there is no evidence that the galaxies in the green valley in our sample are rejuvenating and they are more likely in the process of quenching. 

However, we cannot attribute this quenching to the AGN activity. In Section \ref{sec:sfhs_agns_mass}, we found that the SFRs of the AGN host galaxies are consistent with galaxies at similar masses from the full LEGA-C sample. Since the majority of high mass galaxies of the sample are below the SFMS, we cannot rule out that the suppression of star formation is driven by the higher mass of the AGN sample. Due to AGN variability, the timescale for when AGN are actively accreting and visible is much shorter \citep[$\mathrm{\gtrsim 10^5 yr}$;][]{Hickox2014, Schawinski2015} than even the most rapid quenching events ($\mathrm{\approx 10^8 yr}$). Also, our SFR tracers probe the most recent SFR on timescales of $\mathrm{10^7}$ yrs, which is still longer than the timescale of AGN activity. Figure \ref{fig:mass_sfr_dist} shows that different empirical AGN selections are identifying evidence of AGN with overlapping distributions across the entire sSFR range of the full LEGA-C sample. This means we see evidence for current AGN activity in galaxies above, through, and below the SFMS. This makes it difficult to determine if the instantaneous AGN activity influences the large-scale star formation activity in galaxies. However, there is growing evidence in recent simulations and observational studies that the integrated history of AGN feedback, on Gyr timescales, rather than instantaneous AGN feedback suppresses star formation \citep[][]{Lammers2022, Piotrowska2022, Baker2023, Bluck2023}. \citet{Bluck2023} suggest that AGN release energy over long periods, which prevents gas from cooling and accreting from the CGM into massive galaxies, leading to the quenching of star formation due to the unavailability of fuel. This scenario would explain the difficulty of connecting instantaneous AGN activity with the suppression of star formation in observational studies.

\section{Summary and Conclusions}\label{sec:summary}
We have followed up a subset of LEGA-C galaxies with NIR observations at MMT Observatory, allowing us to select AGN with the [\ion{N}{2}] BPT diagram at higher redshift ($\mathrm{z\sim0.7}$) to look for observational evidence of AGN quenching during the epoch of cosmic star formation decline. We study the nonparametric SFHs of AGN hosts and compare them to star-forming galaxies to check for observational evidence of AGN quenching. Our main findings are as follows:

\begin{itemize}
\item The SFRs inferred from our SED fitting with \texttt{Prospector} are in excellent agreement with the SFRs measured from the dust-corrected $\mathrm{H\alpha}$ line fluxes, showing only 0.1 dex in scatter.

\item $89\pm15\%$ of galaxies in the green valley or below host AGN, based on sSFRs inferred from SED fitting with \texttt{Prospector}. 

\item The AGN host galaxies are $0.37\pm0.15$ dex below the SFMS while the galaxies without detectable AGN are consistent with being on or above the SFMS. However, when comparing to a bootstrapped mass-matched sample, the SFRs of the AGN host galaxies are consistent with the same SFR trends at a given stellar mass with the full LEGA-C sample. Therefore, we cannot rule out that this suppression of star formation is driven by the higher mass of the AGN sample.

\end{itemize}

We therefore show that despite having high quality data and state-of-the-art SFH modeling, it remains difficult to connect the suppression of star formation with current AGN activity. This conclusion should guide the design of future experiments with mass-complete distributions and similar quality data at intermediate redshifts of BPT-selected AGN. This is currently difficult to obtain as it requires large time investments on telescopes in both optical and NIR spectrographs. However, in the near future, surveys using instruments such as the Multi-Object Optical and Near-infrared Spectrograph (MOONS) on the European Southern Observatory (ESO) Very Large Telescope \citep[VLT;][]{Taylor2018}, the spectroscopic galaxy survey with the Dark Energy Spectroscopic Instrument \citep[DESI;][]{DESI2016}, the galaxy evolution survey with the Prime Focus Spectrograph \citep[PFS;][]{Greene2022}, and the wide field redshift surveys planned with Euclid \citep[][]{Euclid2022} will enable large samples with the complete rest-optical emission line diagnostics used in this work. 
This will enable future studies to further investigate the connection between AGN activity and suppressed star formation with larger sample sizes during the peak epoch of quenching.

\begin{acknowledgments}
Observations reported here were obtained at the MMT Observatory, a joint facility of the University of Arizona and the Smithsonian Institution. We thank Sean Moran and Igor Chilingarian for help with the data reduction pipeline and Joannah Hinz for help with the mask design.

Based on observations made with ESO Telescopes at the La Silla Paranal Observatory under programme ID 194-A.2005 (The LEGA-C Public Spectroscopy Survey). CW and REH are supported by the National Science Foundation through the Graduate Research Fellowship Program funded by Grant Award No. DGE-1746060. CCW and MR were supported by the National Aeronautics and Space Administration (NASA) Contract NAS50210 to the University of Arizona.

This material is based upon High Performance Computing (HPC) resources supported by the University of Arizona TRIF, UITS, and Research, Innovation, and Impact (RII) and maintained by the UArizona Research Technologies department.

We respectfully acknowledge the University of Arizona is on the land and territories of Indigenous peoples. Today, Arizona is home to 22 federally recognized tribes, with Tucson being home to the O'odham and the Yaqui. Committed to diversity and inclusion, the University strives to build sustainable relationships with sovereign Native Nations and Indigenous communities through education offerings, partnerships, and community service.
\end{acknowledgments}

\facilities{HST, VLT, MMT (MMIRS)}

\software{\texttt{Prospector} \citep[v1.1.0,][]{Johnson2021}, \texttt{python-fsps} \citep[v0.4.1,][]{python-fsps}, \texttt{sedpy} \citep[v0.3.0,][]{sedpy}, \texttt{fsps} \citep[v3.2,][]{Conroy2009, Conroy2010}, \texttt{astropy} \citep[v5.0.4,][]{astropy:2013, astropy:2018}, \texttt{matplotlib} \citep[v3.5.1,][]{matplotlib}, \texttt{dynesty} \citep[v1.2.3,][]{dynesty:2020}, \texttt{scipy} \citep[v1.7.3][]{scipy}, \texttt{numpy} \citep[v1.22.3,][]{numpy}, \texttt{corner} \citep[v2.2.1,][]{Foreman-Mackey2016}, GELATO \citep[v2.5.1][]{Hviding2022}}

\appendix
\section{Additional Figures}\label{sec:appendix}

\begin{figure*}[!htbp]
\centering
\includegraphics[width=0.7\textwidth]{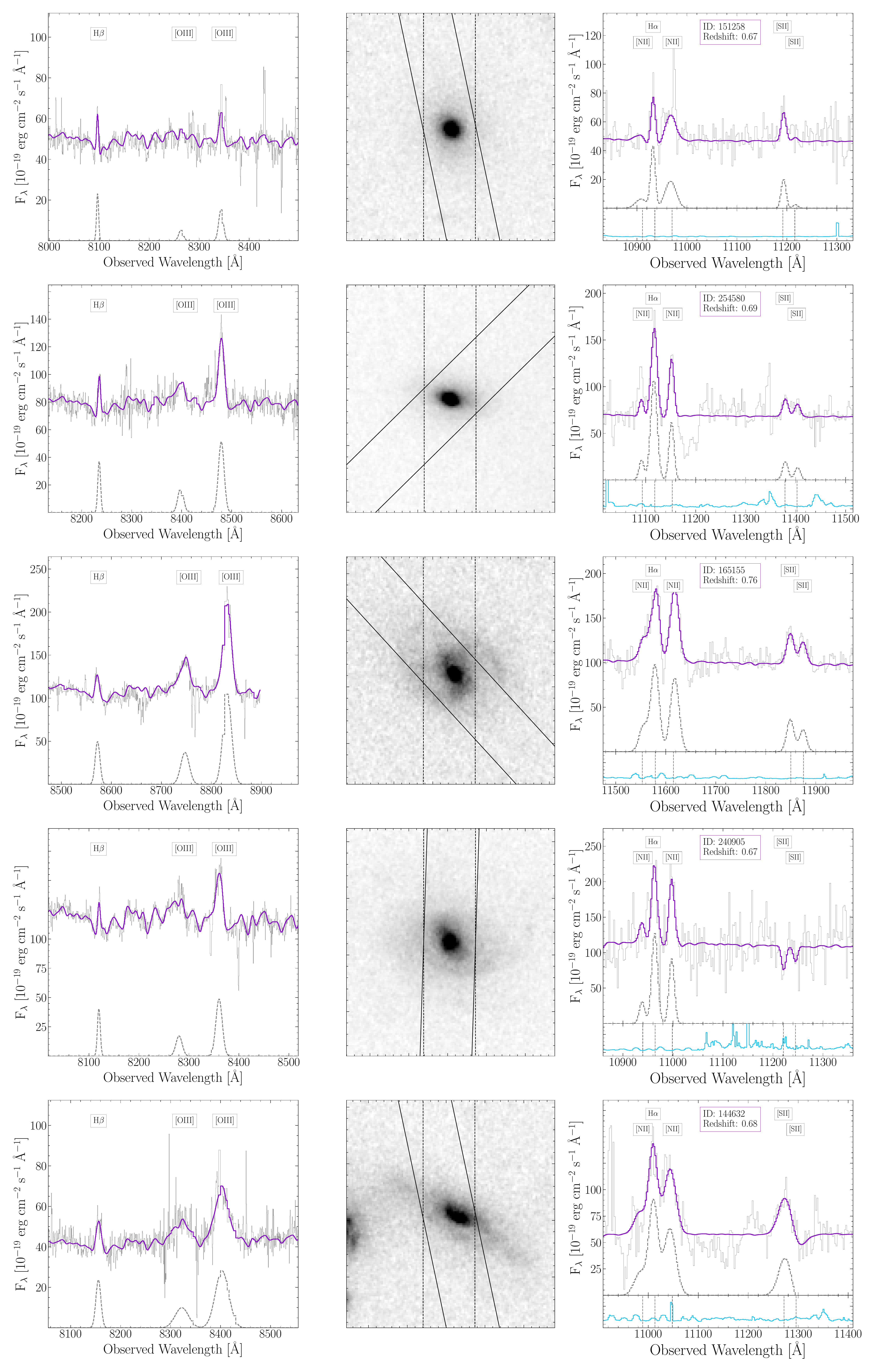}
\caption{This figure shows the observational data for our sample. The GELATO spectroscopic fits are displayed separately for each galaxy, showing the $\mathrm{H\beta}$ and [\ion{O}{3}] emission lines observed by VLT in the leftmost column and the $H\alpha$, [\ion{N}{2}], and [\ion{S}{2}] emission lines observed by MMT in the rightmost column. The subpanels in the rightmost column show the sky lines, obtained from the 2D spectra, as light blue lines. The spectra are plotted as a gray line, the components of the emission lines are plotted as a gray dotted line, and the overall fit is plotted in purple for AGN candidates and yellow for galaxies without detectable AGN. The center column shows the \textit{HST/ACS} F814W images of each galaxy as 4$\times$4 \arcsec cutouts. The slit alignments for the optical LEGA-C observations are shown as dashed lines and the slit alignments for the NIR MMT observations are shown as solid lines. \label{fig:obs1}}
\end{figure*}

\renewcommand{\thefigure}{\arabic{figure} (Cont.)}
\addtocounter{figure}{-1}

\begin{figure*}[!htbp]
\centering
\includegraphics[width=0.7\textwidth]{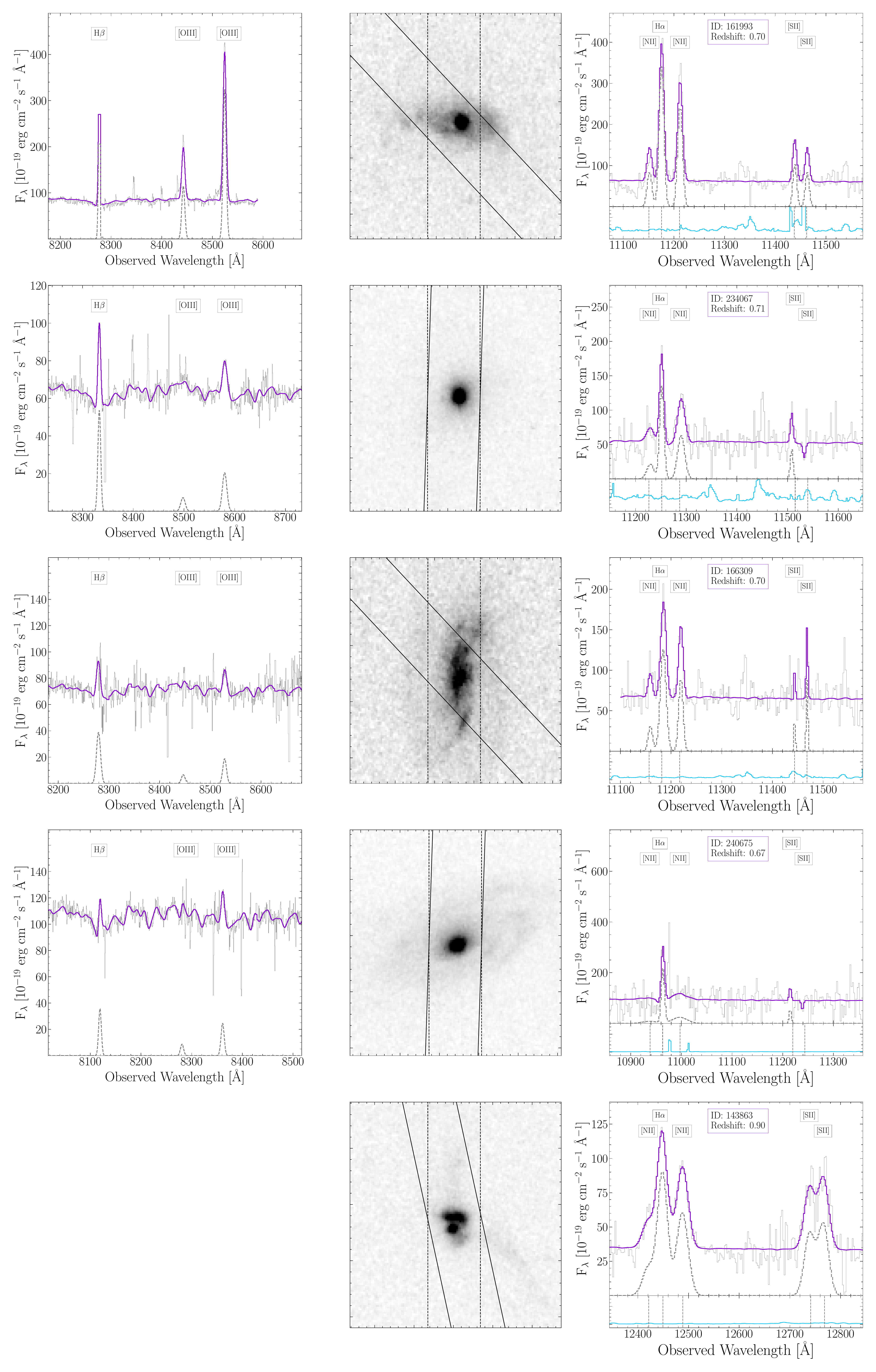}
\caption{ }
\end{figure*}

\renewcommand{\thefigure}{\arabic{figure}}

\renewcommand{\thefigure}{\arabic{figure} (Cont.)}
\addtocounter{figure}{-1}

\begin{figure*}[!htbp]
\centering
\includegraphics[width=0.7\textwidth]{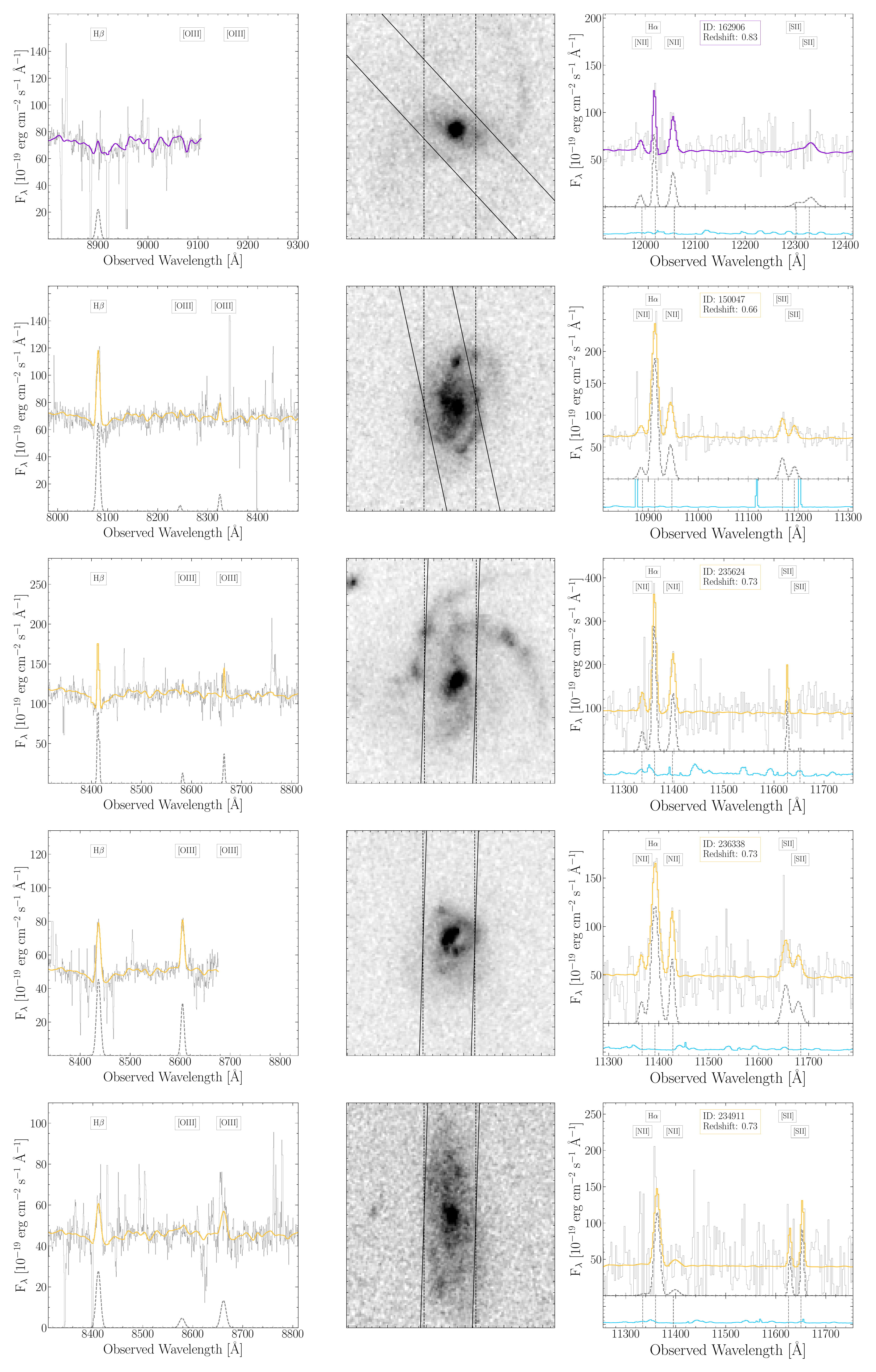}
\caption{ }
\end{figure*}

\renewcommand{\thefigure}{\arabic{figure}}

\renewcommand{\thefigure}{\arabic{figure} (Cont.)}
\addtocounter{figure}{-1}

\begin{figure*}[!htbp]
\centering
\includegraphics[width=0.7\textwidth]{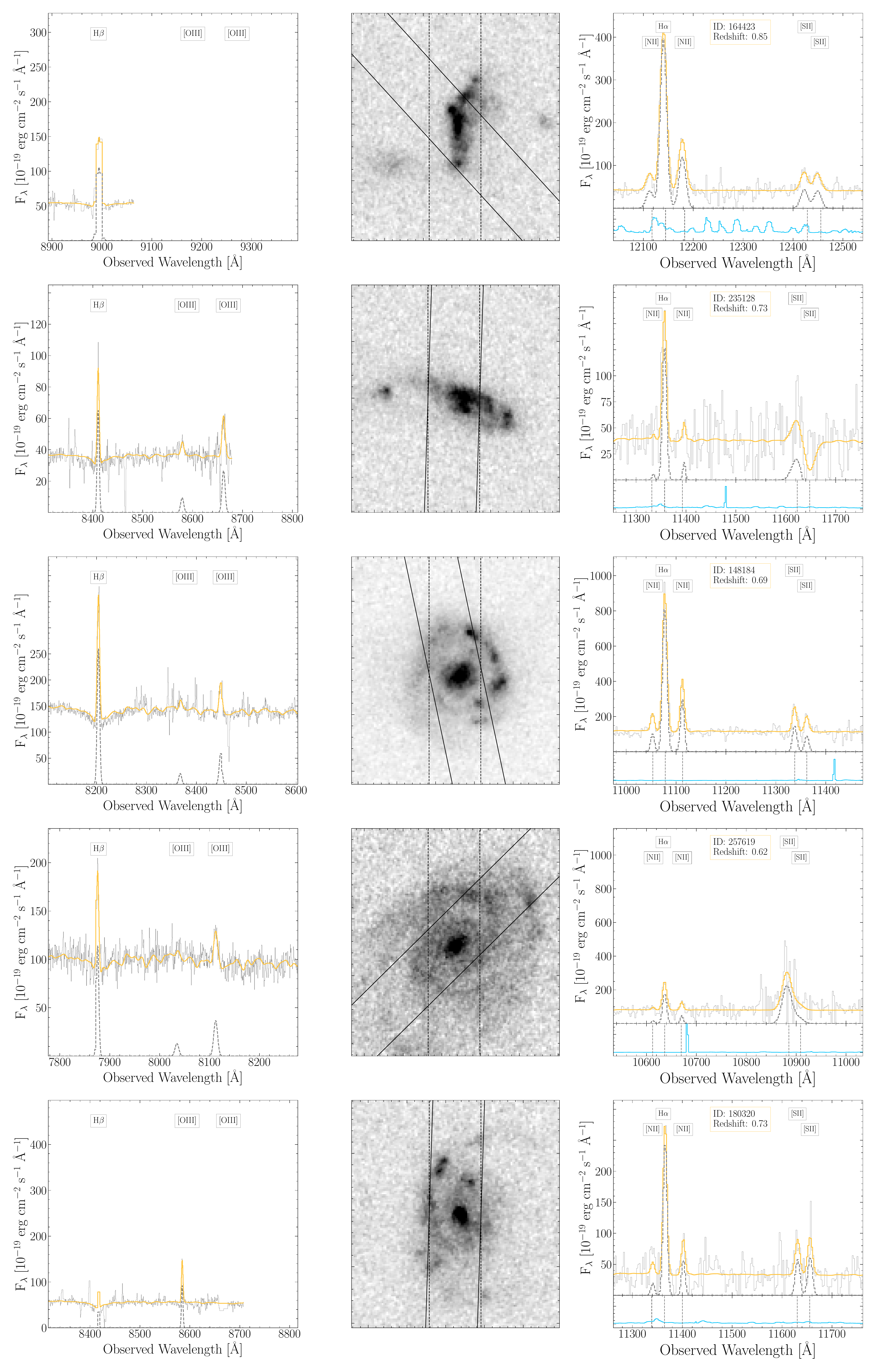}
\caption{ }
\end{figure*}

\renewcommand{\thefigure}{\arabic{figure}}

\renewcommand{\thefigure}{\arabic{figure} (Cont.)}
\addtocounter{figure}{-1}

\begin{figure*}[!htbp]
\centering
\includegraphics[width=0.7\textwidth]{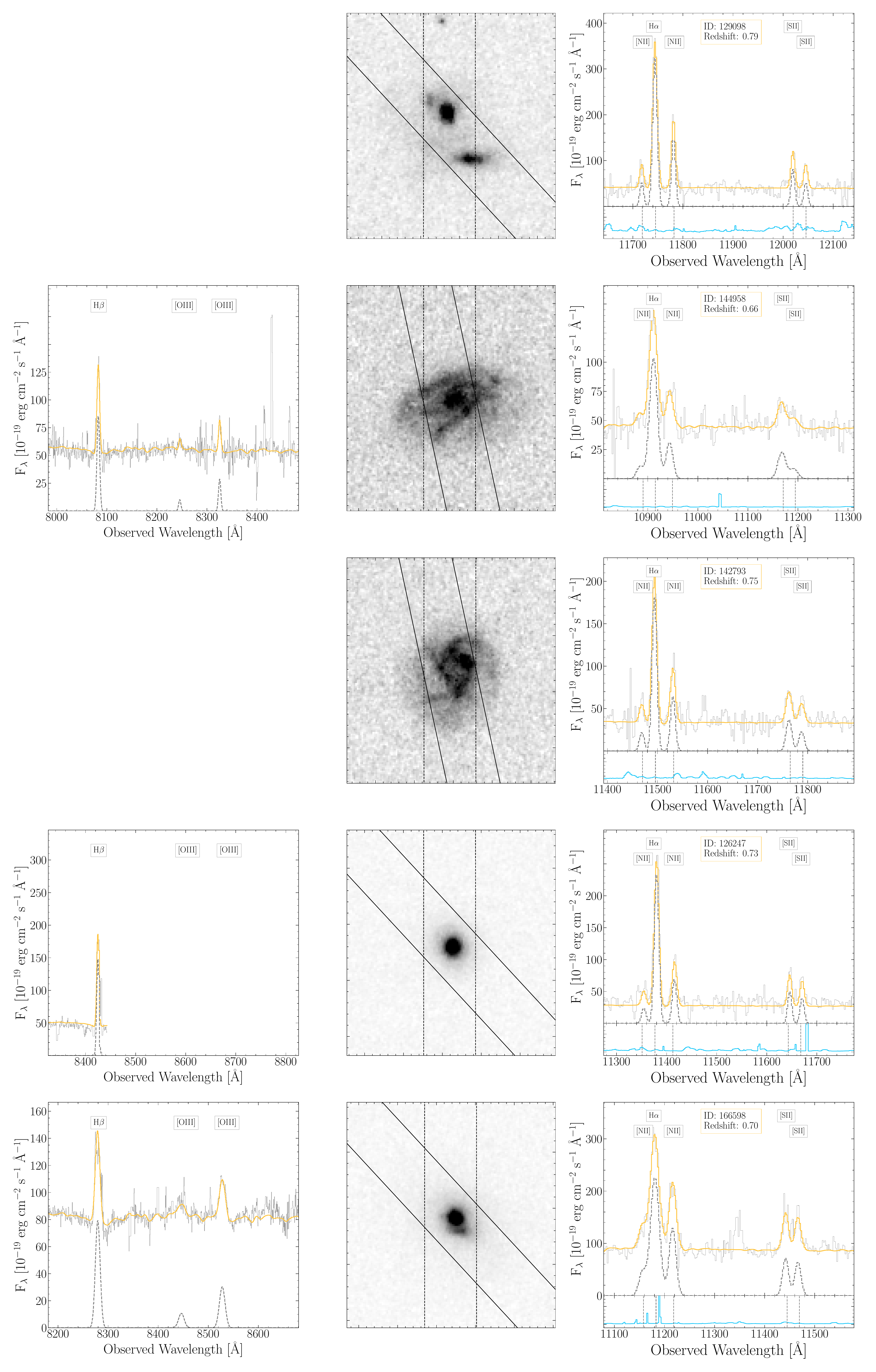}
\caption{ }
\end{figure*}

\renewcommand{\thefigure}{\arabic{figure}}

\renewcommand{\thefigure}{\arabic{figure} (Cont.)}
\addtocounter{figure}{-1}

\begin{figure*}[!htbp]
\centering
\includegraphics[width=0.7\textwidth]{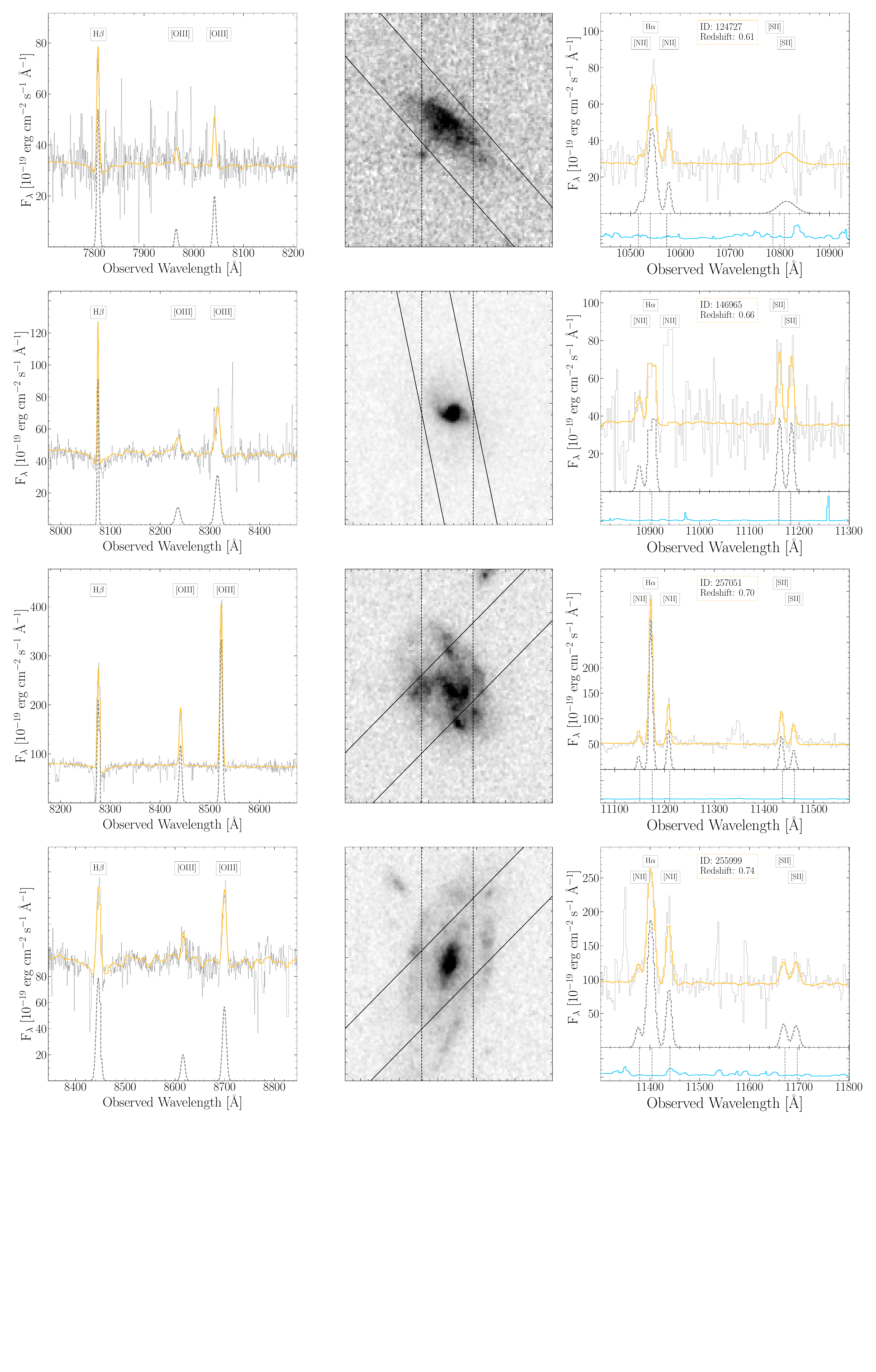}
\caption{ }
\end{figure*}

\renewcommand{\thefigure}{\arabic{figure}}

\begin{rotatetable}
\begin{deluxetable}{rrrrrrrrrrr}
\movetableright=0.05cm
\tablecaption{Measurements \label{tab:measurements}}
\tablehead{
\colhead{\#} & \colhead{ID} & \colhead{$\mathrm{H\beta}$} &  \colhead{[\ion{O}{3}]$\mathrm{\lambda 5007}$\AA} & \colhead{[\ion{N}{2}]$\mathrm{\lambda 6584}$\AA} & \colhead{$\mathrm{H\alpha}$} & \colhead{[\ion{S}{2}] $\mathrm{\lambda 6717}$\AA} & \colhead{[\ion{S}{2}] $\mathrm{\lambda 6731}$\AA} & \colhead{n} & \colhead{$\mathrm{\tau_{dust,1}}$} & \colhead{$\mathrm{\tau_{dust,2}}$}
}
\startdata
1 & 150047 & $ 543.61 \pm 7.64 $ & $ 75.77 \pm 10.89 $ & $ 782.15 \pm 42.35 $ & $ 2802.66 \pm 7.64 $ & $ 433.20 \pm 34.87 $ & $ 263.24 \pm 36.90 $ & $ -0.37 ^{+ 0.07 } _{- 0.08 } $ & $ 0.57 ^{+ 0.24 } _{- 0.04 } $ & $ 0.63 ^{+ 0.04 } _{- 0.04 } $ \\
2 & 235624 & $ 801.64 \pm 55.32 $ & $ 148.02 \pm 16.17 $ & $ 1491.21 \pm 81.58 $ & $ 2846.79 \pm 55.32 $ & $ 517.34 \pm 100.46 $ & $ 39.34 \pm 191.63 $ & $ -0.26 ^{+ 0.05 } _{- 0.05 } $ & $ 0.27 ^{+ 0.11 } _{- 0.10 } $ & $ 0.51 ^{+ 0.04 } _{- 0.04 } $ \\
3 & 240675 & $ 242.11 \pm 8.49 $ & $ 193.47 \pm 17.76 $ & $ 868.93 \pm 218.59 $ & $ 1765.34 \pm 8.49 $ & $ 297.88 \pm 82.24 $ & $ \ldots $ & $ 0.18 ^{+ 0.13 } _{- 0.18 } $ & $ 0.22 ^{+ 0.06 } _{- 0.05 } $ & $ 0.26 ^{+ 0.03 } _{- 0.04 } $ \\
4 & 236338 & $ 476.75 \pm 10.65 $ & $ \ldots $ & $ 752.35 \pm 46.24 $ & $ 2302.21 \pm 10.65 $ & $ 686.07 \pm 88.32 $ & $ 388.44 \pm 47.20 $ & $ -0.17 ^{+ 0.05 } _{- 0.05 } $ & $ 0.93 ^{+ 0.16 } _{- 0.16 } $ & $ 1.17 ^{+ 0.06 } _{- 0.06 } $ \\
5 & 151258 & $ 123.97 \pm 6.38 $ & $ 178.32 \pm 43.96 $ & $ 516.63 \pm 68.21 $ & $ 500.53 \pm 6.38 $ & $ 251.32 \pm 41.61 $ & $ 31.9 \pm 37.63 $ & $ -0.41 ^{+ 0.21 } _{- 0.17 } $ & $ 0.30 ^{+ 0.05 } _{- 0.08 } $ & $ 0.25 ^{+ 0.04 } _{- 0.03 } $ \\
6 & 254580 & $ 258.25 \pm 9.28 $ & $ 741.09 \pm 15.02 $ & $ 778.29 \pm 45.66 $ & $ 1756.37 \pm 9.28 $ & $ 279.19 \pm 42.63 $ & $ 192.76 \pm 33.56 $ & $ -0.00 ^{+ 0.10 } _{- 0.09 } $ & $ 0.39 ^{+ 0.14 } _{- 0.14 } $ & $ 0.62 ^{+ 0.06 } _{- 0.05 } $ \\
7 & 143863 & $ \ldots $ & $ \ldots $ & $ 1722.23 \pm 42.40 $ & $ 2224.60 \pm 36.96 $ & $ 1162.81 \pm 84.06 $ & $ 1351.06 \pm 105.38 $ & $ -0.13 ^{+ 0.02 } _{- 0.00 } $ & $ 0.56 ^{+ 0.04 } _{- 0.15 } $ & $ 0.70 ^{+ 0.02 } _{- 0.00 } $ \\
8 & 234911 & $ 320.27 \pm 7.04 $ & $ 180.63 \pm 15.18 $ & $ 169.97 \pm 102.60 $ & $ 1676.05 \pm 7.04 $ & $ 404.56 \pm 126.28 $ & $ 690.20 \pm 213.27 $ & $ -0.05 ^{+ 0.08 } _{- 0.08 } $ & $ 1.13 ^{+ 0.19 } _{- 0.18 } $ & $ 0.81 ^{+ 0.05 } _{- 0.05 } $ \\
9 & 164423 & $ 1165.39 \pm 28.53 $ & $ \ldots $ & $ 2096.75 \pm 44.87 $ & $ 6437.72 \pm 28.53 $ & $ 796.57 \pm 62.52 $ & $ 740.77 \pm 32.91 $ & $ -0.07 ^{+ 0.04 } _{- 0.04 } $ & $ 0.18 ^{+ 0.12 } _{- 0.10 } $ & $ 0.93 ^{+ 0.06 } _{- 0.06 } $ \\
10 & 235128 & $ 345.70 \pm 8.10 $ & $ 207.10 \pm 14.64 $ & $ 116.02 \pm 28.09 $ & $ 1342.14 \pm 8.10 $ & $ 567.75 \pm 149.20 $ & $ \ldots $ & $ 0.02 ^{+ 0.03 } _{- 0.03 } $ & $ 0.15 ^{+ 0.17 } _{- 0.10 } $ & $ 1.39 ^{+ 0.06 } _{- 0.06 } $ \\
11 & 148184 & $ 1626.36 \pm 11.57 $ & $ 426.33 \pm 19.99 $ & $ 2850.63 \pm 44.02 $ & $ 9151.36 \pm 11.57 $ & $ 1593.18 \pm 92.59 $ & $ 967.74 \pm 62.25 $ & $ -0.19 ^{+ 0.04 } _{- 0.04 } $ & $ 0.02 ^{+ 0.03 } _{- 0.01 } $ & $ 0.55 ^{+ 0.03 } _{- 0.03 } $ \\
12 & 257619 & $ 676.78 \pm 45.86 $ & $ 356.46 \pm 19.11 $ & $ 376.27 \pm 48.65 $ & $ 1879.07 \pm 45.86 $ & $ 5271.99 \pm 417.27 $ & $ 823.77 \pm 145.40 $ & $ 0.12 ^{+ 0.20 } _{- 0.21 } $ & $ 0.37 ^{+ 0.07 } _{- 0.06 } $ & $ 0.32 ^{+ 0.05 } _{- 0.06 } $ \\
13 & 180320 & $ 1537.02 \pm 971.65 $ & $ 1023.87 \pm 18.30 $ & $ 517.99 \pm 48.51 $ & $ 2433.68 \pm 971.65 $ & $ 683.81 \pm 105.30 $ & $ 729.14 \pm 102.26 $ & $ 0.23 ^{+ 0.09 } _{- 0.10 } $ & $ 0.18 ^{+ 0.10 } _{- 0.10 } $ & $ 0.33 ^{+ 0.06 } _{- 0.06 } $ \\
14 & 129098 & $ \ldots $ & $ \ldots $ & $ 1506.81 \pm 58.63 $ & $ 3741.03 \pm 56.02 $ & $ 810.42 \pm 55.85 $ & $ 517.37 \pm 40.53 $ & $ -0.45 ^{+ 0.06 } _{- 0.05 } $ & $ 0.86 ^{+ 0.20 } _{- 0.16 } $ & $ 1.03 ^{+ 0.07 } _{- 0.06 } $ \\
15 & 165155 & $ 591.55 \pm 10.66 $ & $ 2261.70 \pm 23.44 $ & $ 1819.91 \pm 64.54 $ & $ 2030.13 \pm 10.66 $ & $ 659.01 \pm 40.90 $ & $ 448.01 \pm 41.96 $ & $ 0.17 ^{+ 0.11 } _{- 0.11 } $ & $ 0.62 ^{+ 0.09 } _{- 0.07 } $ & $ 0.47 ^{+ 0.05 } _{- 0.05 } $ \\
16 & 162906 & $ 221.65 \pm 19.67 $ & $ \ldots $ & $ 443.28 \pm 55.28 $ & $ 638.55 \pm 19.67 $ & $ 105.16 \pm 142.37 $ & $ 227.18 \pm 121.23 $ & $ -0.28 ^{+ 0.49 } _{- 0.54 } $ & $ 0.07 ^{+ 0.01 } _{- 0.01 } $ & $ 0.05 ^{+ 0.05 } _{- 0.03 } $ \\
17 & 144958 & $ 586.93 \pm 11.60 $ & $ 170.06 \pm 8.69 $ & $ 528.66 \pm 35.52 $ & $ 2071.99 \pm 11.60 $ & $ 457.35 \pm 91.78 $ & $ 170.44 \pm 118.27 $ & $ -0.06 ^{+ 0.08 } _{- 0.07 } $ & $ 0.58 ^{+ 0.16 } _{- 0.15 } $ & $ 0.66 ^{+ 0.05 } _{- 0.05 } $ \\
18 & 142793 & $ \ldots $ & $ \ldots $ & $ 783.22 \pm 46.73 $ & $ 2179.31 \pm 43.10 $ & $ 526.15 \pm 35.32 $ & $ 325.37 \pm 43.33 $ & $ -0.06 ^{+ 0.06 } _{- 0.08 } $ & $ 0.44 ^{+ 0.31 } _{- 0.09 } $ & $ 0.59 ^{+ 0.07 } _{- 0.06 } $ \\
19 & 126247 & $ 858.59 \pm 14.78 $ & $ \ldots $ & $ 798.58 \pm 27.64 $ & $ 2782.05 \pm 14.78 $ & $ 511.91 \pm 40.85 $ & $ 412.43 \pm 24.56 $ & $ -0.27 ^{+ 0.05 } _{- 0.04 } $ & $ 0.04 ^{+ 0.05 } _{- 0.03 } $ & $ 0.40 ^{+ 0.04 } _{- 0.02 } $ \\
20 & 166598 & $ 948.38 \pm 14.91 $ & $ 432.64 \pm 8.73 $ & $ 2254.69 \pm 68.40 $ & $ 5109.72 \pm 14.91 $ & $ 1109.67 \pm 98.01 $ & $ 991.29 \pm 59.74 $ & $ -0.11 ^{+ 0.06 } _{- 0.05 } $ & $ 0.93 ^{+ 0.16 } _{- 0.14 } $ & $ 0.93 ^{+ 0.05 } _{- 0.04 } $ \\
21 & 124727 & $ 345.20 \pm 9.21 $ & $ 143.29 \pm 8.15 $ & $ 209.51 \pm 30.72 $ & $ 936.47 \pm 9.21 $ & $\ldots$ & $ 308.94 \pm 86.79 $ & $ -0.02 ^{+ 0.05 } _{- 0.05 } $ & $ 0.79 ^{+ 0.23 } _{- 0.41 } $ & $ 1.00 ^{+ 0.08 } _{- 0.06 } $ \\
22 & 240905 & $ 289.93 \pm 9.38 $ & $ 777.85 \pm 16.98 $ & $ 1244.6 \pm 128.93 $ & $ 1913.73 \pm 9.38 $ & $\ldots$ & $\ldots$ & $ -0.24 ^{+ 0.30 } _{- 0.28 } $ & $ 0.07 ^{+ 0.02 } _{- 0.02 } $ & $ 0.07 ^{+ 0.03 } _{- 0.02 } $ \\
23 & 234067 & $ 380.32 \pm 6.61 $ & $ 226.81 \pm 8.70 $ & $ 1112.90 \pm 71.72 $ & $ 1471.69 \pm 6.61 $ & $ 292.13 \pm 54.74 $ & $ \ldots $ & $ 0.02 ^{+ 0.06 } _{- 0.06 } $ & $ 0.77 ^{+ 0.24 } _{- 0.16 } $ & $ 0.81 ^{+ 0.04 } _{- 0.05 } $ \\
24 & 166309 & $ 389.88 \pm 16.67 $ & $ 170.60 \pm 10.00 $ & $ 984.33 \pm 82.98 $ & $ 1811.97 \pm 16.67 $ & $ 140.21 \pm 137.52 $ & $ 371.74 \pm 80.64 $ & $ -0.20 ^{+ 0.08 } _{- 0.08 } $ & $ 0.59 ^{+ 0.18 } _{- 0.13 } $ & $ 0.62 ^{+ 0.04 } _{- 0.04 } $ \\
25 & 146965 & $ 338.46 \pm 4.72 $ & $ 376.19 \pm 8.85 $ & $ \ldots $ & $ 1252.12 \pm 4.72 $ & $ 425.75 \pm 67.01 $ & $ 400.84 \pm 42.15 $ & $ -0.17 ^{+ 0.05 } _{- 0.06 } $ & $ 0.42 ^{+ 0.15 } _{- 0.24 } $ & $ 0.72 ^{+ 0.05 } _{- 0.05 } $ \\
26 & 257051 & $ 1126.76 \pm 12.70 $ & $ 2184.56 \pm 8.24 $ & $ 657.26 \pm 13.46 $ & $ 2706.12 \pm 12.70 $ & $ 628.64 \pm 36.02 $ & $ 371.64 \pm 19.50 $ & $ -0.01 ^{+ 0.07 } _{- 0.07 } $ & $ 0.13 ^{+ 0.09 } _{- 0.08 } $ & $ 0.45 ^{+ 0.06 } _{- 0.05 } $ \\
27 & 255999 & $ 804.54 \pm 13.06 $ & $ 557.00 \pm 10.15 $ & $ 1159.54 \pm 127.44 $ & $ 3251.99 \pm 13.06 $ & $ 545.97 \pm 37.65 $ & $ 506.54 \pm 52.00 $ & $ -0.12 ^{+ 0.05 } _{- 0.06 } $ & $ 0.73 ^{+ 0.27 } _{- 0.22 } $ & $ 0.96 ^{+ 0.05 } _{- 0.06 } $ \\
28 & 161993 & $ 1864.95 \pm 119.22 $ & $ 2514.25 \pm 10.84 $ & $ 2986.60 \pm 67.86 $ & $ 4057.83 \pm 119.22 $ & $ 1025.88 \pm 89.62 $ & $ 823.83 \pm 66.01 $ & $ 0.22 ^{+ 0.07 } _{- 0.07 } $ & $ 0.36 ^{+ 0.10 } _{- 0.11 } $ & $ 0.60 ^{+ 0.05 } _{- 0.05 } $ \\
29 & 144632 & $ 298.85 \pm 12.77 $ & $ 950.91 \pm 34.98 $ & $ 1902.20 \pm 90.73 $ & $ 1967.41 \pm 12.77 $ & $ 1293.00 \pm 95.30 $ & $ \ldots $ & $ 0.02 ^{+ 0.04 } _{- 0.05 } $ & $ 1.10 ^{+ 0.18 } _{- 0.21 } $ & $ 1.45 ^{+ 0.06 } _{- 0.06 } $ \\
\enddata
\tablecomments{Columns 3-8 are flux measurements from the GELATO fitting results in units of $\mathrm{10^{-19}\ erg\ cm^{-2}\ s^{-1}\ \text{\AA}^{-1}}$. Columns 9-11 are the dust measurements from the \texttt{Prospector} fits and includes all components of the dust model: power-law modifier to the shape of the attenuation curve of the diffuse dust ($n$), the birth cloud dust ($\mathrm{\tau_{dust,1}}$), and the diffuse dust screen ($\mathrm{\tau_{dust,2}}$). The [\ion{S}{2}] line fluxes are fitted completely independently. The [\ion{N}{2}]$\mathrm{\lambda 6584}$\AA line is fit independently and the [\ion{N}{2}]$\mathrm{\lambda 6548}$\AA line is fixed to have 0.34/1 times its flux.}
\end{deluxetable}
\end{rotatetable}
\newpage
\bibliography{main}{}
\bibliographystyle{aasjournal}

\end{document}